\documentclass[breakmath]{jsedi_pub}
\usepackage[titletoc]{appendix}
\usepackage{chngcntr}
\usepackage{amsmath}
\usepackage{xfrac}
\usepackage{xcolor}
\DeclareMathOperator{\atantwo}{atan2}

\definecolor{lightgray}{gray}{0.5}

\let\oldv\verbatim
\let\oldendv\endverbatim

\def\verbatim{\par\setbox0\vbox\bgroup\oldv}
\def\endverbatim{\oldendv\egroup\fboxsep0pt \noindent\colorbox[gray]{0.8}{\usebox0}\par}

\title{Joint Bayesian inference of Earth’s magnetic field and core surface flow on millennial timescales}
\shorttitle{Joint inference of core field and surface flow}

\author[1]{Andreas~Nilsson
	\orcid{0000-0002-9528-1276}
	\thanks{Corresponding author: 
\href{mailto:andreas.nilsson@mgeo.lu.se}{\texttt{andreas.nilsson@mgeo.lu.se}}}
}
\author[1]{Neil~Suttie
	\orcid{0000-0001-7374-5423}
}
\author[1]{Marie~Troyano
}
\author[2]{Nicolas~Gillet
	\orcid{0000-0002-2219-1026}
}
\author[3]{Julien~Aubert
	\orcid{0000-0002-2756-0724}
}
\author[1]{Anders~Irbäck
	\orcid{0000-0003-1384-0626}
}
\affil[1]{Department of Earth and Environmental Sciences, Lund University, Lund, Sweden}
\affil[2]{Univ. Grenoble Alpes, Univ. Savoie Mont Blanc, CNRS, IRD, Univ. Gustave Eiffel, Grenoble, France}
\affil[3]{Institut de Physique du Globe de Paris, CNRS, Universit\'{e} Paris Cite, Paris, France}

\credit{Conceptualization}{Andreas Nilsson, Neil Suttie, Nicolas Gillet}
\credit{Formal Analysis}{Andreas Nilsson, Neil Suttie, Marie Troyano}
\credit{Writing - Original draft}{Andreas Nilsson}
\credit{Writing - Review \& Editing}{Andreas~Nilsson, Neil~Suttie, Marie~Troyano, Nicolas~Gillet, Julien~Aubert, Anders~Irbäck}

\begin{document}

\publicationonly{
\dois{10.46298/jsedi.17320}
\handedname{Alexandre Fournier}
\receiveddate{January 16, 2026}
\reviseddate{March 19, 2026}
\accepteddate{March 26, 2026}
\publisheddate{March 31, 2026}
\theyear{2026}
\thevolume{2}
\thepaper{1}  
}

\makesedititle{
  \begin{summary}{Abstract}
    Understanding Earth’s core dynamics over millennial timescales requires models that jointly describe the evolution of the geomagnetic field and core surface flow, while accommodating the sparse, irregular, and uncertain nature of archaeomagnetic and palaeomagnetic data. We present a new Bayesian core field and core flow modelling framework that utilises archaeo/palaeomagnetic data directly, combining a reduced stochastic representation of core surface dynamics derived from numerical geodynamo statistics with a probabilistic treatment of observational and chronological uncertainties. A key innovation is an efficient discrete marginalisation of age uncertainties, which avoids the convergence difficulties associated with co-estimating ages in high-dimensional Hamiltonian Monte Carlo inversions. The framework aims to reconstruct the coupled evolution of the geomagnetic field and core surface flow over the past 9000 years while preserving dynamical correlations implied by the prior geodynamo time series. Tests using synthetic data generated from an Earth-like geodynamo demonstrate that the method reliably recovers large-scale geomagnetic field variations and key aspects of core dynamics, including long-term westward drift and the evolution of planetary-scale eccentric gyres. These results show that, when combined with physically informed priors, archaeo/palaeomagnetic data can constrain millennial-scale core flow, paving the way for reconstructions based on real data.
  \end{summary}
  \begin{summary}{Non-technical summary}
    Earth’s magnetic field is generated by the motion of its liquid iron-rich core. Here, we introduce a new modelling approach that uses magnetic measurements from archaeological materials and geological records to estimate how both Earth’s magnetic field and the flow at the top of the core have evolved over the past 9000 years. The method combines information from physics-based simulations of Earth’s core with a statistical framework that accounts for measurement errors and dating uncertainties. Tests using artificial data show that the method can recover large-scale changes in the magnetic field and key patterns of core motion, such as persistent westward drift and planetary-scale gyres. These results demonstrate that archaeological and geological magnetic records contain enough information to reveal long-term core dynamics, opening the door to future reconstructions using real data.
\end{summary}
}

\section{Introduction}

The past two decades of satellite monitoring of the magnetic field, in combination with advancements in numerical simulations of the geodynamo, have generated a wealth of knowledge of processes in Earth's core on centennial to annual timescales \citep[e.g.][]{Gillet2022, Aubert2023}. Geodynamo simulations have been able to reproduce the main characteristics of the Earth’s magnetic field and core surface flow over the era of direct observations \citep[e.g.][]{Aubert2013,Mound2023}, e.g. the preference for strong secular variation in the Atlantic hemisphere \citep{Finlay2020}. Comparatively less attention has been concentrated on reproducing characteristic features of geomagnetic field variations over the past $\sim$10,000 years, mostly due to the lack of reference data with adequate resolution and disagreement between palaeomagnetic models \citep{Wardinski2025}. On these timescales, spanning several convective overturn times ($\sim$130 years), external forcing mechanisms and boundary conditions such as lower mantle heat-flux heterogeneities \citep[e.g.][]{Mound2023}, gravitational coupling between inner-core and mantle and asymmetric inner core growth \citep[e.g.][]{Aubert2013,Aubert2017}, are thought to play a more dominant role.

On millennial timescales, the observational constraints required to investigate core dynamics are provided by indirect records of the geomagnetic field. Through recent technical innovations, models based on these archaeo/palaeomagnetic observations of Earth’s magnetic field are providing information on past field changes with unprecedented resolution \citep[e.g.][]{Nilsson2014,Constable2016,Hellio2018,Campuzano2019,Nilsson2022,Schanner2022}. Characteristic field observations, including patterns of westward/eastward drift and recurrent field asymmetries, have tentatively been interpreted in terms of large-scale flow \citep[e.g.][]{Dumberry2007,Nilsson2020,Nilsson2022}. However, due to the many challenges involved, so far only a handful of attempts have been made to actually reconstruct core surface flow based on these field models \citep[e.g.][]{Dumberry2006,Wardinski2008a,Kianishahvandi2024,Suttie2025,Rivera2026}.

One of the most recent efforts by \textcite{Suttie2025}, takes advantage of statistics taken from numerical dynamo simulations \citep{Aubert2021} to address key challenges with the contributions of sub-grid errors and magnetic diffusion to induction at the core surface, which can be significant even on short timescales with high-degree field models. The study, which is based on the \texttt{pygeodyn} geomagnetic data assimilation package \citep{Gillet2019,Huder2019,istas2023}, demonstrates that despite the low spatial and temporal resolution attainable with archaeo/palaeomagnetic data, it is possible to reproduce the large-scale features of modern core surface flow reconstructions over the past century \citep{Gillet2019} as well as changes in length-of-day derived from historical eclipses over the past 1200 years \citep{Morrison2001}. Key ingredients for this are the physical constraints provided by the geodynamo simulations. Although not conclusive, the study also provides evidence that the persistent westward drift of 0.09°/year previously observed in palaeomagnetic field models \citep{Nilsson2020, Nilsson2022} is tied to the long-term average zonal flow in Earth's core.

A common feature of all previous efforts to model the core surface flow on millennial timescales is that they use the Gauss coefficients of existing palaeomagnetic field models as input data, i.e. as observations. This is not ideal given that variations in core flow are primarily dependent on secular variation that is poorly constrained in these models and often highly influenced by the choice of modelling methodology. In addition, the uneven geographical distribution of the data, which are mostly concentrated in the northern hemisphere, may introduce systematic errors in the models \citep{Nilsson2024} that are in turn not properly addressed in the core flow reconstruction.

Modelling geomagnetic field variations (and potentially also core surface flow) through archaeo/palaeomagnetic observations involves major challenges related to chronological data uncertainties and post-depositional remanent magnetisations (pDRM). The latter affects sedimentary records and leads to an unknown smoothing and temporal offset of the recorded signal \citep{Roberts2004,Mellstrom2015,Nilsson2018,Bohsung2023}. \textcite{Nilsson2021} introduced a probabilistic method that addresses these challenges by co-estimating the ages of the archaeo/palaeomagnetic data as model parameters and forward modelling the effects of pDRM, using an efficient Hamiltonian Monte Carlo (HMC) algorithm \citep{Carpenter2017} to estimate the parameters. The main bottleneck of this modelling strategy is the co-estimation of ages, in effect synchronizing the data using the geomagnetic field. As demonstrated by \textcite{Nilsson2021}, data with large age uncertainties, in the range of the variability of the geomagnetic field, can lead to multimodal posteriors (bimodal marginal distributions) that are difficult for the HMC algorithm to handle. The method also introduces a lot of parameters, one for each of the $N=9440$ archaeomagnetic observations \citep{Nilsson2022}, which is particularly problematic when applied to core flow models that already occupy comparatively large parameter spaces.

Here we present a new method to model the core magnetic field and core surface flow directly constrained by archaeo/palaeomagnetic data. The method combines the stochastic representation of core surface dynamics from \texttt{pygeodyn} \citep{Huder2019}, derived using statistics from geodynamo simulations, with the probabilistic method to incorporate archaeo/palaeomagnetic observations of \textcite{Nilsson2021}. To address the limitations related to chronological data uncertainties we have introduced an efficient new approach to discretely marginalize the age parameters. The model approach is tested using synthetic data generated from an Earth-like dynamo simulation to evaluate the spatial and temporal range of core surface flow that can feasibly be recovered using the proposed methodology.
\section{Archaeo/palaeomagnetic data}
\label{section_data}

Since the aim is to evaluate a new method to model core surface flow, we will only consider synthetic data in this study. To build this dataset we start from the archaeo/palaeomagnetic data used to construct the pfm9k.2 model \citep{Nilsson2022}, including all the pre-treatments implemented for that model. The dataset consists of $N_{ARC}=9440$ archaeomagnetic observations and $N_{SED}=1465$ sedimentary observations from $N_{S}=10$ records covering the past 9000 years. To generate synthetic data we use the Midpath Coupled-Earth dynamo simulation \citep{Aubert2017} and extract a 10,000 year period to represent our reference geomagnetic field and core surface flow over the period 8000 BCE to 2000 CE. We choose this longer reference time series to be able to generate synthetic data with "true" ages older than 7000 BCE. The choice of the Midpath simulation is motivated by its Earth-like properties, with the ability to resolve geophysical variations on timescales of $\sim$20 years (the Alfvén time in the simulation) and longer, which is enough for the purpose of modelling the core field and flow evolution from archaeo-/palaeomagnetic data. Compared to more recent simulations \citep{Aubert2021,Aubert2023}, Midpath has also been run over a comparatively long integration time ($\sim$20,000 years), which will be important for later aspects of the model construction.

Using the reference geomagnetic field, the synthetic dataset was constructed with the identical spatial and temporal distribution as the real data following a similar procedure as described in \textcite{Nilsson2021}. All synthetic observations of declination, $D$, inclination, $I$ and intensity, $F$, were generated at the same locations as in the real data but using randomly generated reference ages consistent with the age uncertainties associated with the respective data. The archaeomagnetic $D,I,F$ observations were then further corrupted with noise, consistent with the given data uncertainties, drawn from a Student's t distribution with $\nu=4$ degrees of freedom to simulate the presence of outliers. The sediment $D,I$ observations (no sedimentary intensity data were included) were filtered and offset to account for the effects of pDRM, core orientation and inclination shallowing errors, with the associated reference parameters drawn from the respective prior distributions (see section \ref{section_SED_data_model}) and then corrupted with Gaussian noise based on the data uncertainties.
\section{Core field and flow model}

Our core field and flow (CFF) model is constructed following a similar approach used to build the pfm9k.2 geomagnetic field model \citep{Nilsson2021,Nilsson2022}, but with two key modifications: (i) the Gaussian Process based geomagnetic field parameterisation is replaced with a discrete-time stochastic representation of the core surface field and flow \citep{Gillet2019,Huder2019} and (ii) the (parameterised) ages of the archaeo/palaeomagnetic observations are discretely integrated out (marginalised).

The model follows a Bayesian approach, where the prior distribution of the core flow and field, informed by a dynamo simulation statistics, is updated using archaeo/palaeomagnetic observations via the likelihood function. An ensemble of models, i.e. samples from the joint posterior distribution, are generated using a HMC algorithm, specifically the No-U-Turn sampler implemented through Stan 2.34.1 \citep{Carpenter2017}. An important difference to existing Kalman filter approaches \citep[e.g.][]{Gillet2019,Suttie2025}, is that the models are not sequentially built up through time with forecast model trajectories iteratively adjusted as new observations are added. Instead, in our model, the entire time series is constructed as one long uninterrupted series of forecast steps that is evaluated against available observations simultaneously in the likelihood function. This ensures that all spatial and temporal cross correlations defined in the prior statistics, based on the dynamo simulation, are preserved and obeyed to the extent that they are consistent with the observations.

For the model description we use the following notation style: For any vectors $\mathbf{x}(t)$ and $\mathbf{y}(t)$ the time average is $\mathbf{x}_0=E(\mathbf{x}(t))$, which unless otherwise stated is calculated from a dynamo simulation time series. The perturbation to the average is $\mathbf{x'}(t) = \mathbf{x}(t) - \mathbf{x}_0$ and the cross-covariance matrix $\textsf{K}_{xy}=E(\mathbf{x'}(t) \, \mathbf{y'}(t)^T)$. The corresponding correlation matrix is $\textsf{R}_{xy} = \text{diag}(\textsf{K}_{xx})^{-1/2}\, \textsf{K}_{xy}\,  \text{diag}(\textsf{K}_{yy})^{-1/2}$,
with $\text{diag}(\textsf{K}_{xx})$ the diagonal matrix whose diagonal elements are identical to those of $K_{xx}$. Empirically derived cross-covariances obtained from a dynamo series $\mathbf{x}^*(t)$ and $\mathbf{y}^*(t)$ are denoted $\textsf{K}^*_{xy}$.

\begin{figure*}[t]
\centering
\includegraphics[width=0.8\textwidth]{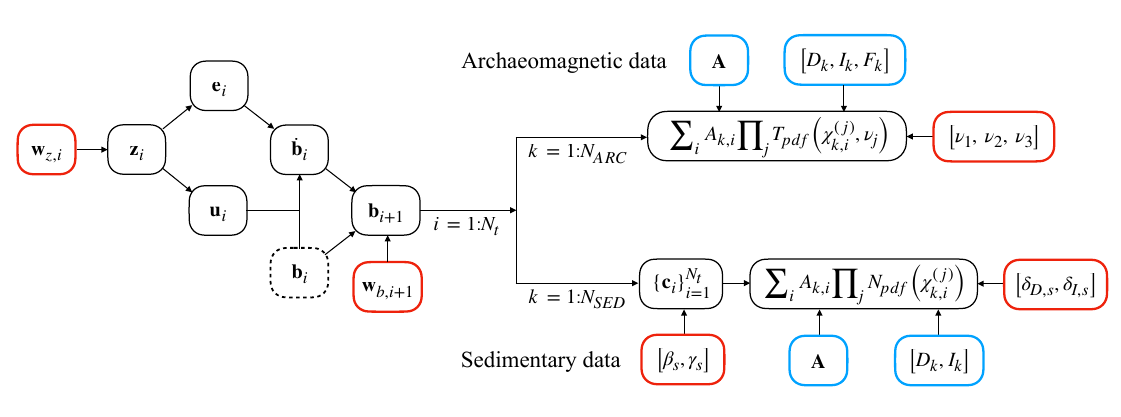}
\caption{A schematic chart of the model with data highlighted in blue, model parameters (variables) in red and transformed parameters/quantities in black (for more details see main text). The left-hand part of the chart depicts the iterative construction of the core field and core flow discrete time series, at times $t_1,...,t_{N_t}$. The right-hand part of the chart depicts the likelihood function which is split into two blocks of data: Archaeomagnetic data ($k = 1, \dots, N_{ARC}$) and Sedimentary data ($k = 1, \dots, N_{SED}$, from $s = 1, \dots, N_{S}$ records).}
\label{Fig_schematic}
\end{figure*}

Fig. \ref{Fig_schematic} shows a schematic overview of the model with detailed descriptions of each part provided in the following three sections.

\subsection{Parameterisation of core field and flow}

The temporal evolution of the radial magnetic field is described by the radial part of the induction equation at the core-mantle boundary (CMB)
\begin{equation}
    \label{induction_eq}
    \frac{\partial B_r}{\partial t} = -\nabla_h \cdot (\mathbf{U}_h B_r) + \frac{\eta}{r} \nabla^2 (rB_r),
\end{equation}
with $B_r$ the radial magnetic field at the CMB, $\mathbf{U}_h$ the core surface flow, $\nabla_h \cdot$ the horizontal part of the divergence operator and $\eta$ the magnetic diffusivity. The first term on the right-hand side (r.h.s.) represents the advection of the radial field and second term its diffusion. $B_r$ and $\mathbf{U}_h$ are parameterised in the spectral domain based on spherical harmonic (SH) expansions.

We adopt a spherical coordinate system $(r;\theta;\phi)$, where $r$ is radius, $\theta$ is colatitude and $\phi$ is longitude. We take the Earth's radius as $r_e = 6371.2$ km and core radius as $r_c = 3485$ km. At $r\ge r_c$, assuming an insulating mantle, the core magnetic field $\mathbf{B}$ can be described as the gradient of a scalar potential, $\mathbf{B} = -\nabla V$, with
\begin{equation}
    \label{potential}
    V(r,\theta,\phi) = r_e\sum_{l=1}^{L_b} \sum_{m=0}^l \left(\frac{r_e}{r}\right)^{l+1} [g_l^m \cos{m\phi}+h_l^m\sin{m\phi}] P_l^m(\cos{\theta}),
\end{equation}
where $g_l^m$ and $h_l^m$ are the Gauss coefficients,  $P_l^m(\cos{\theta})$ the associated Schmidt semi-normalised Legendre functions of degree $l$ and order $m$ and $L_b$ the truncation degree. Based on this SH representation it is straightforward to obtain field components $B_r, B_{\theta}, B_{\phi}$ and in turn $D,I,F$ for any given location ($r \ge r_c,\theta,\phi$), see \textcite{Langel1987}. The main field (MF) Gauss coefficients and their time derivatives, i.e. the secular variation (SV) coefficients, are stored in vectors $\mathbf{b}$ and $\dot{\mathbf{b}}$ of dimensions $N_b=L_b(L_b+2)$. 

The core surface flow is represented as \citep[e.g.][]{Holme2015}
\begin{equation}
    \label{Uh}
    \mathbf{U}_h(\theta, \phi) = \nabla \times (T\mathbf{r}) + \nabla_h(rS),
\end{equation}
with $T$ and $S$ respectively the toroidal and poloidal scalars. $T$ and $S$ are also expanded in the basis of Schmidt semi-normalised SH:
\begin{align}
    & T(\theta,\phi) = \sum_{l=1}^{L_u} \sum_{m=0}^l [t_l^{m,c} \cos{m\phi} + t_l^{m,s}\sin{m\phi}] P_l^m(\cos{\theta}),
    \label{T_scalars}
    \\
    & S(\theta,\phi) = \sum_{l=1}^{L_u} \sum_{m=0}^l [s_l^{m,c} \cos{m\phi} + s_l^{m,s}\sin{m\phi}] P_l^m(\cos{\theta}),
    \label{S_scalars}
\end{align}
with $t_l^{m,c}, t_l^{m,s}$ and $s_l^{m,c}, s_l^{m,s}$ the toroidal and poloidal flow coefficients and $L_u$ the flow truncation degree. The toroidal and poloidal flow coefficients are stored in vector $\mathbf{u}$ of dimension $N_u=2L_u(L_u+2)$.

Following \textcite{Gillet2019}, we introduce a decomposition of the radial induction equation \eqref{induction_eq} into the resolvable large-scale processes and the unresolved sub-grid processes and diffusion, which can be written in the spectral domain as
\begin{align}
    \label{induction_eq_spectral_domain}
    \dot{\mathbf{b}} = \textsf{A}(\mathbf{b})\mathbf{u} + \mathbf{e} = \mathbf{f} + \mathbf{e}.
\end{align}
The first term on the r.h.s., $\mathbf{f}=\textsf{A}(\mathbf{b})\mathbf{u}$, corresponds to the SV induced by large-scale flow where the matrix $\textsf{A}$ accounts for the Gaunt and Elsasser integrals \citep{Whaler1986} (as well as the downward continuation of $\mathbf{b}$ to the CMB). The second term on the r.h.s., $\mathbf{e}$, is the error of representativeness, which accounts for the contribution to SV arising from both sub-grid processes and diffusion. The vectors $\mathbf{f}$ and $\mathbf{e}$, of dimension $N_b$, contains SH coefficients $\{f_l^{m,c},f_l^{m,s}\}$ and $\{e_l^{m,c},e_l^{m,s}\}$ that are defined equivalently to the MF and SV coefficients through eq. \eqref{potential}. 

Given the inherent limitations of the archaeo/palaeomagnetic data, the large scale (modelled) MF and SV are defined by a truncation at SH degree $L_b=5$ \citep{Licht2013,Sanchez2016,Nilsson2022}, while the modelled core flow is limited to SH degree $L_u=10$ \citep{Suttie2025}. We emphasize that the higher degree flows are retained to ensure that the modelled core flows are consistent with the types of flow seen in the dynamo simulations, but this does not necessarily mean that they can be resolved. The spatial complexity of the flow is further limited by only using the first $N_v$ principal components. The principal component analysis (PCA) is based on a time series of the core surface coefficients from the geodynamo simulation and $N_v=46$ (in the case of Midpath) set to retain $95\%$ of the variance. The flow vector can then be written as
\begin{equation}
    \label{reduced_rank}
    \mathbf{u}=\mathbf{u}_0 + \Phi\mathbf{v},
\end{equation}
with $\mathbf{u}_0$ defined as the time-averaged flow from the geodynamo simulation, $\Phi$ the projection operator and $\mathbf{v}$ the reduced rank flow vector.

Contrary to most other models of core flow \citep[e.g.][]{Pais2008,Gillet2019}, ours corresponds to a pure forward model with no need to actually calculate the full $\textsf{A}(\mathbf{b})$ matrix, let alone its inverse form. A considerable speed-up in computation time, which is essential for the HMC implementation, is achieved by instead calculating the whole term $\textsf{A}(\mathbf{b})\mathbf{u}$ directly using products of the magnetic field and the flow in the spatial domain. We use a Gauss-Legendre grid with 16 nodes in $\theta$ (the zeros of the degree-16 Legendre polynomial) and 32 uniformly spaced points in $\phi$. The induced SV is then back-projected onto the spherical harmonics basis using a Gauss-Legendre Quadrature method. This is similar to how the term $\textsf{A}(\mathbf{b})$ is usually calculated in practice \citep{Lloyd1990}, but requires a factor of $2L_u$ less calculations. The main benefit, however, is that the design matrices required to go from spectral to spatial domain for a pre-defined grid, can all be pre-calculated and passed along to the HMC algorithm.

\subsection{Multivariate core field and flow process}

We concatenate $\mathbf{v}$ with $\mathbf{e'}$ to form an augmented core state vector $\mathbf{z}^T = [\mathbf{v}^T \mathbf{e'}^T]$, leaving aside the time-averaged error of representativeness $\mathbf{e}_0$ considered known from the dynamo simulation. The temporal evolution of $\mathbf{z}(t)$ and $\mathbf{b}(t)$ are modelled using a time-discrete stochastic process in increments of $\Delta t$ over $N_t$ time steps $(t_1,...,t_{N_t})$. The statistical properties of the stochastic process are derived from a prior geodynamo simulation sampled at $\Delta t^* = \Delta t = 50$ years (in the case of Midpath dynamo series). 

In the following, we describe how the time-discrete vectors $\mathbf{z}_i$ and $\mathbf{b}_i$ at times $t_i$ are constructed, parameterised entirely by the 'white noise' terms $\mathbf{w}_{z,i}$ and $\mathbf{w}_{b,i}$ with independent, standard normal priors (table \ref{table_prior}).

\begin{table}[b] 
\centering
\begin{tabular}{l l}
\hline
$\forall i\in[1,...,N_t]$ & $\mathbf{w}_{z,i} \sim \mathcal{N}(0,\textsf{I}_z)$ \\
& $\mathbf{w}_{b,i} \sim \mathcal{N}(0,\textsf{I}_b)$ \\
[1ex] 
$\forall j\in[1,2,3]$ & $(\nu_j-1) \sim \Gamma(2,0.1)$ \\
[1ex] 
$\forall s\in[1,...,N_s]$ & $\beta_s \sim \Gamma(3,2)$ \\
& $\gamma_s \sim \Gamma(2,0.2)$ \\
& $\delta D_s \sim \mathcal{N}(0,5^2)$ \\
& $\delta I_s \sim \mathcal{N}(0,1)$ \\
\hline
\end{tabular}
\caption{Prior distributions for model parameters, where $\textsf{I}_z$ and $\textsf{I}_b$ are identity matrices of rank $N_z$ and $N_b$ respectively and $\Gamma(\cdot)$ denotes the Gamma distribution. See \textcite{Nilsson2021} and \textcite{Nilsson2022} for justification of the choice of priors for $\nu_j, \beta_s, \gamma_s, \delta D_s$ and $\delta I_s$.}
\label{table_prior}
\end{table}

The time series are initialized by generating a random core state $\mathbf{z}_1$ and core field $\mathbf{b}_1$ with spatial cross-covariances $\textsf{K}_{zz} = E(\mathbf{z} \mathbf{z}^T)$ and $\textsf{K}_{bb} = E(\mathbf{b'} \mathbf{b'}^T)$ designed to match those of the prior geodynamo time series $\mathbf{z}^*(t)$ and $\mathbf{b}^*(t)$, i.e $\textsf{K}_{zz}^*$ and $\textsf{K}_{bb}^*$:
\begin{align}
    & \mathbf{z}_1 = \textsf{L}_z\mathbf{w}_{z,1},
    \\
    & \mathbf{b}_1 = \textsf{L}_b\mathbf{w}_{b,1}+\mathbf{b}_0,
\end{align}
where $\textsf{L}_z {\textsf{L}_z}^T = \textsf{K}_{zz}$ and $\textsf{L}_b {\textsf{L}_b}^T = \textsf{K}_{bb}$. In other words, $\textsf{L}_z$ and $\textsf{L}_b$ are the Cholesky decompositions of $\textsf{K}_{zz}$ and $\textsf{K}_{bb}$. For any given time (in this case $t_i=t_1$), the core state vector $\mathbf{z}_i$ gives access to $\mathbf{u}_i$ and $\mathbf{e}_i$, which combined with $\mathbf{b}_i$ are used to obtain $\dot{\mathbf{b}}_i$ through eq. \eqref{induction_eq_spectral_domain}.

Numerical integration of the core state vector, $\mathbf{z}_{i+1}$, is based on a multivariate AR-1 process, discretized using an Euler–Maruyama scheme \citep{Gillet2019}
\begin{equation}
    \label{z_i+1}
    \mathbf{z}_{i+1} = (\textsf{I}_z - \Delta t \textsf{D}_z) \, \mathbf{z}_i + \sqrt{\Delta t} \, \mathbf{r}_i,
\end{equation}
where $\textsf{I}_z$ is the identity matrix of rank $N_z$ and $\textsf{D}_z$ the drift matrix. The vector $\mathbf{r}_i$ represents Gaussian noise, uncorrelated in time and with cross-covariance $\textsf{K}_{rr} = E(\mathbf{r}_i {\mathbf{r}_i}^T)$ constructed such that the recovered $\mathbf{z}(t)$ attains the appropriate cross-covariance $\textsf{K}_{zz}$ (see below). To calculate $\mathbf{r}_i$ we take the Cholesky decomposition $\textsf{L}_r$ of $\textsf{K}_{rr}$ to obtain $\mathbf{r}_i = \textsf{L}_r \mathbf{w}_{z,i+1}$.

\begin{figure*}[t]
\centering
\includegraphics[width=1.0\textwidth]{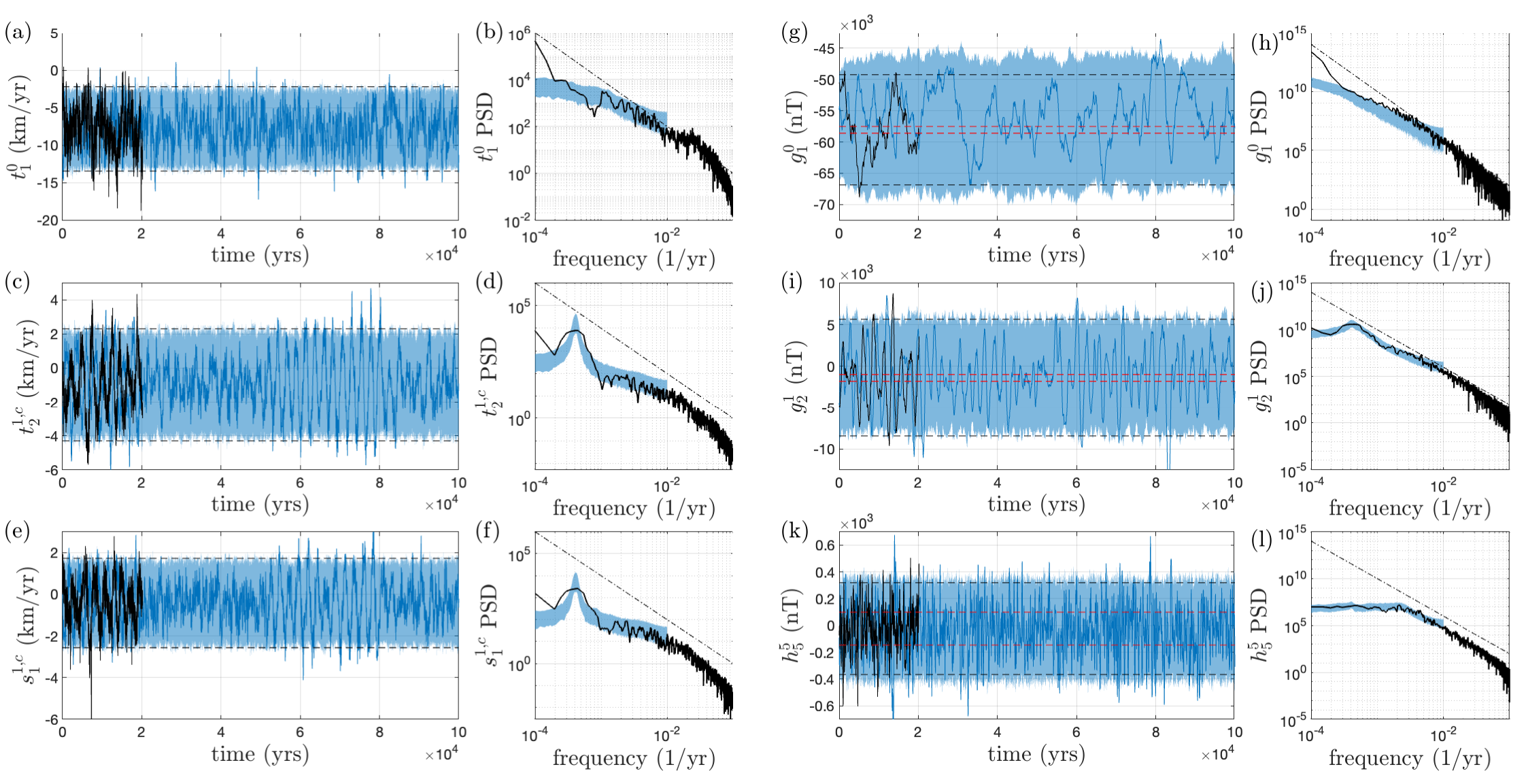}
\caption{Time series and temporal power spectra of (a-f) three core flow coefficients and (g-l) three MF gauss coefficients. The Midpath prior geodynamo series (black) is compared to a 100k long random time series generated with the multivariate core flow and field process (blue), with the shaded blue areas representing the 95\% range of 300 random samples. Dashed black lines show the 95\% interval ($\pm 2\sigma$) of the prior dynamo series and the dashed red line shows the equivalent range from the 'white noise' term in eq. \eqref{b_i+1}. Black dot-dashed lines corresponds to a spectral index of $p = 2$ and $p = 4$, (i.e. $P(f) \propto f^{-p}$) for core flow and core field coefficients respectively. Note that the differences in the power spectra towards long periods are due to the limited duration for the Midpath dynamo series.}
\label{Fig_forecast}
\end{figure*}

To obtain $\textsf{K}_{rr}$ and the drift matrix $\textsf{D}_z$, corresponding to a ‘restoring force’ in eq. \eqref{z_i+1}, we first define the following cross covariance matrices:
\begin{align}
    \textsf{K}_{zz^+} = & E(\mathbf{z}_i \mathbf{z}_{i+1}^T),
    \\
    \textsf{K}_{\Delta z \Delta z} = & E(\Delta \mathbf{z} \Delta \mathbf{z}^T),
\end{align}
where $\Delta \mathbf{z} = \mathbf{z}_{i+1} - \mathbf{z}_i$. We note that in the following where we define the same cross covariance matrices for the prior geodynamo series, a more general form replacing $\mathbf{z}_i$ and $\mathbf{z}_{i+1}$ with $\mathbf{z}(t)$ and $\mathbf{z}(t + \Delta t)$, respectively, would have been more appropriate. However, since $\Delta t$ is the same for the process sampled by eq. \eqref{z_i+1} and the prior geodynamo series, we keep the subscript notation for brevity. Deriving $\textsf{K}_{zz^+}$ from eq. \eqref{z_i+1}, noting that $E(\mathbf{z}_i\, \mathbf{r}_i^T) = 0$, we get
\begin{equation}
    \textsf{D}_z = \frac{\textsf{I}_z - ({\textsf{K}_{zz}}^{-1} \textsf{K}_{zz^+})^T}{\Delta t}.
\end{equation}
Equivalently deriving $\textsf{K}_{\Delta z \Delta z}$ from eq. \eqref{z_i+1} we get
\begin{equation}
    \textsf{K}_{rr} = \frac{\textsf{K}_{\Delta z \Delta z}}{\Delta t} - \Delta t \textsf{D}_z \textsf{K}_{zz} {\textsf{D}_z}^T.
\end{equation}
$\textsf{D}_z$ and $\textsf{K}_{rr}$ are then estimated from the prior geodynamo time series as $\textsf{D}_z^*$ and $\textsf{K}_{rr}^*$ using a least squares algorithm \citep{Neumaier2001,Gillet2019}. In Fig. \ref{Fig_forecast}a-f we compare time series and temporal spectra of individual
core flow coefficients for the stochastic model series and the prior geodynamo series. The comparison shows, as previously demonstrated \citep{Gillet2019}, that the stochastic model is able to adequately reproduce the spatial and temporal core flow variability of the prior geodynamo time series.

\begin{figure*}[t]
\centering
\includegraphics[width=1.0\textwidth]{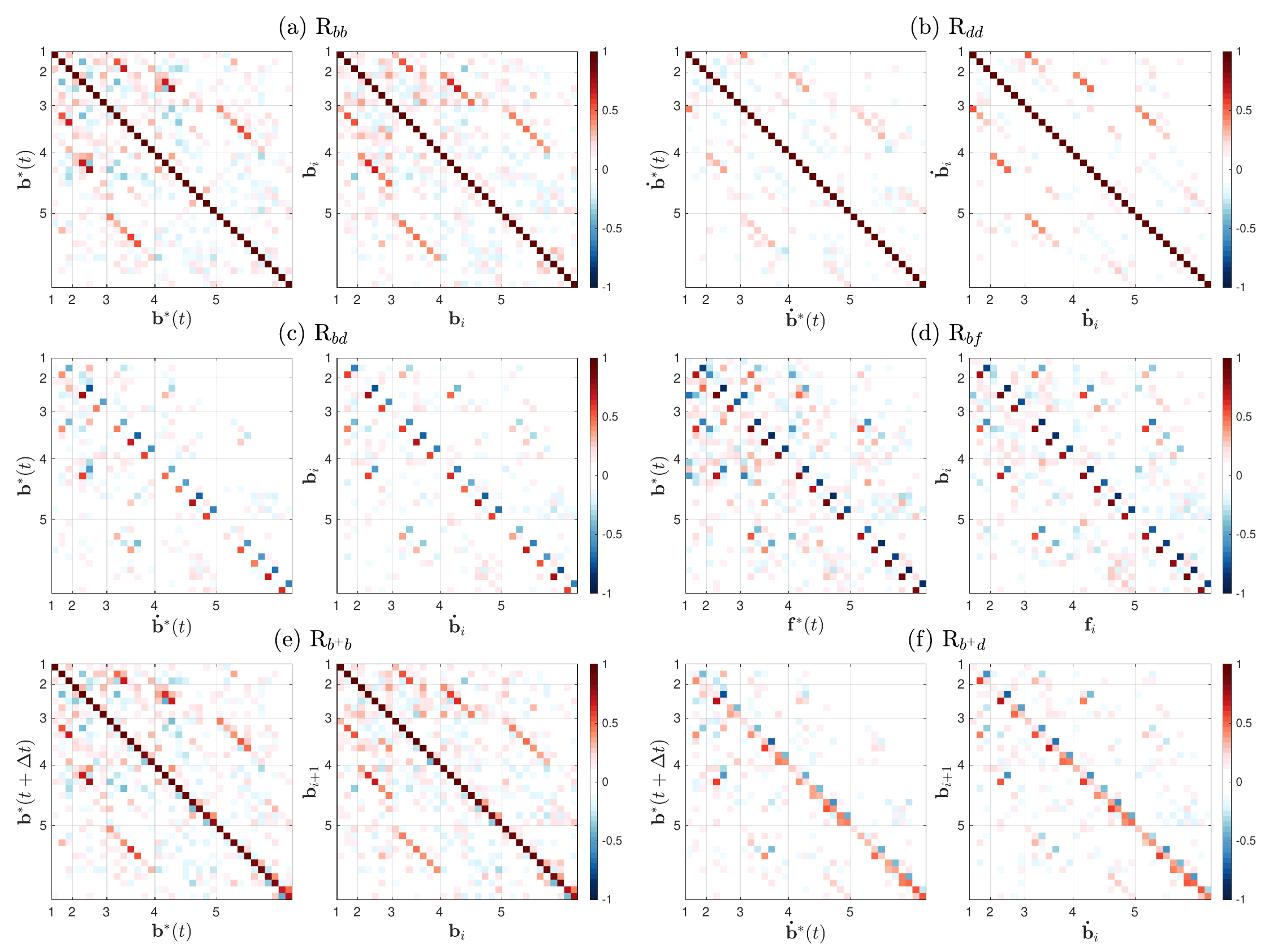}
\caption{Comparison of cross-correlation matrices for various MF and SV vectors (see eq. \eqref{cov_matrices}) empirically derived from the Midpath prior geodynamo series (left hand plot) and a 100k long random time series generated with the multivariate core field and flow process (right hand plot). For each plot the Gauss coefficients are ordered according to the following sequence: $[g_1^0,g_1^1,h_1^1,g_2^0,g_2^1,h_2^1,g_2^2,h_2^2,...,g_5^5,h_5^5]$}
\label{Fig_corr_matrices}
\end{figure*}

The forward propagation of the MF vector, $\mathbf{b}_{i+1}$, differs from the Euler scheme used in the \texttt{pygeodyn} Kalman filter forecast steps \citep{Huder2019,Gillet2019}, where $\Delta t$ are generally small and changes in $\dot{\mathbf{b}}$ can be considered negligible. For large $\Delta t$ (50-100 years), a simple forward propagation based on the instantaneous SV,  $\dot{\mathbf{b}}_i$, does not work, since we also have to account for changes in both the MF and SV in between $t_i$ and $t_{i+1}$. Instead, we formulate the forward propagation of the MF vector as a Gaussian process regression \citep{Rasmussen2006} conditioned on MF and SV observations at $t_i$, $\mathbf{x}_i^T=[\mathbf{b'}_i^T \, \dot{\mathbf{b}}_i^T]$:
\begin{equation}
    \label{D_b x}
    E(\mathbf{b'}_{i+1} \, | \, \mathbf{x}_i) = 
    \textsf{K}_{b^+x} {\textsf{K}_{xx}}^{-1} \mathbf{x}_i
     \equiv \textsf{D}_b\mathbf{x}_i,
\end{equation}
\begin{equation}
     \label{K_bb|x}
     E(\mathbf{b'}_{i+1} \, {\mathbf{b'}_{i+1}}^T \, | \, \mathbf{x}_i) = 
    \textsf{K}_{bb}-\textsf{K}_{b^+x} {\textsf{K}_{xx}}^{-1} {\textsf{K}_{b^+x}}^T
    \equiv \textsf{K}_{bb|x}.
\end{equation}
The cross-covariance matrices $\textsf{K}_{xx}$ and $\textsf{K}_{b^+x}$ that constitute the main building blocks in eqs. \eqref{D_b x} and \eqref{K_bb|x}, are defined as
\begin{align}
    \textsf{K}_{xx} = &
    \begin{bmatrix}
    \textsf{K}_{bb} & \textsf{K}_{bd} \\
    {\textsf{K}_{bd}}^T & \textsf{K}_{dd}
    \end{bmatrix},
    \\
    \textsf{K}_{b^+x} = &
    \begin{bmatrix}
    \textsf{K}_{b^+b} & \textsf{K}_{b^+d} \\
    \end{bmatrix},
\end{align}
with
\begin{align}
    \label{cov_matrices}
    \textsf{K}_{bd} = & E\left( \mathbf{b}'_i \, \dot{\mathbf{b}}_i^T \right),
    \\
    \textsf{K}_{dd} = & E\left( \dot{\mathbf{b}}_i \, \dot{\mathbf{b}}_i^T \right),
    \\
    \textsf{K}_{b^+b} = & E\left( \mathbf{b}'_{i+1} \, {\mathbf{b}'_i}^T \right),
    \\
    \textsf{K}_{b^+d} = & E\left( \mathbf{b}'_{i+1} \, \dot{\mathbf{b}}_i^T \right),
\end{align}
obtained empirically from the prior geodynamo time series as $\textsf{K}_{bd}^*$, $\textsf{K}_{dd}^*$, $\textsf{K}_{b^+b}^*$ $\textsf{K}_{b^+d}^*$ (Fig. \ref{Fig_corr_matrices}). The numerical integration of the MF vector can then be written as
\begin{equation}
    \label{b_i+1}
    \mathbf{b}_{i+1} = \mathbf{b}_0 + \textsf{D}_b \mathbf{x}_i + L_{bx}\mathbf{w}_{b,i+1},
\end{equation}
where $\textsf{L}_{bx} {\textsf{L}_{bx}}^T = \textsf{K}_{bb|x}$. The second term on the r.h.s. provides the link to $\dot{\mathbf{b}}_i$ (through $\mathbf{x}_i$) that in turn is related to the core state vector $\mathbf{z}_i$ ($\mathbf{u}_i$ and $\mathbf{e}_i$) through eq. \eqref{induction_eq_spectral_domain}. Combined with the first term, and for small $\Delta t$, this is effectively equivalent to an Euler scheme, whereas for large $\Delta t$ it converges to the time-average $\mathbf{b}_0$. The third 'white-noise' term on the r.h.s is necessary to account for unmodelled changes in MF and SV and to obtain a stationary distribution for $\mathbf{b}(t)$ (Fig. \ref{Fig_forecast}g-l). For large $\Delta t$, the independently parameterised 'white noise' term will dominate, which means that the recovered variations in  $\mathbf{b}(t)$ are partially decoupled from $\mathbf{z}(t)$. Tests based on the Midpath dynamo series, with various $\Delta t$, shows that the size white noise term is inversely related to MF correlation time $\tau_b=\sqrt{\sigma_b^2/\sigma_{\dot{b}}^2}$, where $\sigma_b$ and $\sigma_{\dot{b}}$ are the square-root diagonal elements of $K_{bb}$ and $K_{dd}$ respectively. As demonstrated in Fig. \ref{Fig_forecast}, the white noise term is mainly important for the higher order and degree Gauss coefficients with short correlation times compared to $\Delta t=50$ years (Fig. \ref{Fig_forecast}k), whilst the large-scale and slowly varying magnetic field components (Fig. \ref{Fig_forecast}g,j) remain primarily the result of core dynamics captured by $\mathbf{z}(t)$. As a rule of thumb, we find that for $\Delta t=\tau_b$, the white noise term added in \eqref{b_i+1} corresponds to roughly 20\% of the total MF variance. In other words, the choice of $\Delta t$ determines the degree to which the modelled core surface flow dynamics are linked to the MF variations that we can observe (section \ref{section_data_model}).

Fig. \ref{Fig_corr_matrices} shows cross-correlation matrices derived empirically from the prior dynamo simulation and from random 100k long (discrete) time series of $\mathbf{z}(t)$ and $\mathbf{b}(t)$ (as shown in Fig. \ref{Fig_forecast}) generated using the multivariate core field and flow process, according to eq. \eqref{z_i+1} and \eqref{b_i+1}. As the comparison demonstrates, the main cross-correlations between core field and flow are adequately preserved in our prior stochastic representation. We note particularly strong positive cross-correlations between $h_l^m$ and $\dot{g}_l^m$ as well as negative cross-correlations between $g_l^m$ and $\dot{h}_l^m$ for $m\neq 0$ (Fig. \ref{Fig_corr_matrices}c). The correlations can be traced to the induced part of the SV (Fig. \ref{Fig_corr_matrices}d) and are related to the predominant westward zonal flow of the prior dynamo simulation, advecting flux in a clockwise direction. 

\subsection{Data model}
\label{section_data_model}

The model likelihood is split into two parts that links the discrete time series $\mathbf{b}(t)$ to the archaeomagnetic and sedimentary palaeomagnetic data, respectively. The methodology closely follows that of \textcite{Nilsson2021}, but introduces a new approach to address chronologic data uncertainties. To account for the error introduced by the SH truncation at $L_b=5$, additional uncertainties of $\sigma_{IT}=1.4^{\circ}$ for inclinations, $\sigma_{DT}=\sigma_{IT}/\cos{I}$ for declinations and $\sigma_{FT}=2\,\mu$T for intensities are added in quadrature to all the data uncertainties \citep{Nilsson2021}.

\subsubsection{Archaeomagnetic data}
To simplify notation we assume that each archaeomagnetic observation $\mathbf{o}_k$ contains all three elements $\{D_k,I_k,F_k\}$, with associated uncertainties $\{\sigma_{D,k},\sigma_{I,k},\sigma_{F,k}\}$. However, in the actual dataset it is often the case that one or two elements are missing. The age of each magnetic field observation is treated as a parameter and assigned a normal prior distribution, defined by the age estimate provided in the literature $a_k \sim \mathcal{N}(\mu_{a,k},{\sigma_{a,k}}^2)$. Contrary to \textcite{Nilsson2021}, who co-estimate the archaeomagnetic ages with the geomagnetic field, we use a discrete method to integrate out, or marginalise, the age parameters. First, the prior probabilities of $a_k$ are calculated discretely for each model time step $t_i$:
\begin{equation}
    A_{k,i} = \int_{t_i-\Delta t/2}^{t_i+\Delta t/2} \frac{1}{\sigma_{a,k}\sqrt{2\pi}} \exp\left(-\frac{(t-\mu_{a,k})^2}{2{\sigma_{a,k}}^2}\right) dt.
\end{equation}
To account for impossible future ages, the probability distributions are truncated at year 2000 CE and each row of the matrix $\mathbf{A}$ is re-normalised appropriately. We note that there are no archaeomagnetic data in the dataset with known ages post 2000 CE \citep{Brown2015a}. Next, the normalised model-data residuals of $\{D,I,F\}$ are similarly calculated for each model time step $t_i$
\begin{align}
    & \chi_{k,i}^{(1)} = \frac{D_k - \textsf{H}_{k,D}(\mathbf{b}_i)}{\sigma_{D,k}},
    \label{chi_D}
    \\
    & \chi_{k,i}^{(2)} = \frac{I_k - \textsf{H}_{k,I}(\mathbf{b}_i)}{\sigma_{I,k}},
    \label{chi_I}
    \\
    & \chi_{k,i}^{(3)} = \frac{F_k - \textsf{H}_{k,F}(\mathbf{b}_i)}{\sigma_{F,k}},
\end{align}
where ${\textsf{H}_{k,D},\textsf{H}_{k,I},\textsf{H}_{k,F}}$ are non-linear forward operators based on eq. \eqref{potential}. The likelihood is then defined as the weighted sum of the probabilities of the magnetic field observation $\mathbf{o}_k$ given $\mathbf{b}(t)$, calculated for each model time step $(t_1,...,t_N)$
\begin{equation}
    \label{ARC_likelihood}
    P(\mathbf{o}_k\, |\, \mathbf{b}_1,...,\mathbf{b}_N,\mathbf{A},\boldsymbol{\nu}) = \sum_{i=1}^{N_t} A_{k,i} \prod_{j=1}^3
    T_{pdf} \left( \chi_{k,i}^{(j)},\nu_j \right),
\end{equation}
where $T_{pdf} \left( \chi,\nu\right)$ is the Student's t probability of $\chi$ with $\nu$ degrees of freedom. Note that, common to previous modelling efforts, the errors in the different field components $(D,I,F)$ are assumed to be independent, which is a reasonable assumption as long as the errors are small. The Student's t error distribution is introduced to account for systematic uncertainties, i.e. outlier data. Following \textcite{Nilsson2021}, we assign Gamma priors (table \ref{table_prior}) for the number of degrees of freedom parameters $\boldsymbol{\nu}$, one for each magnetic field component ($D,I,F$). In practice, time steps associated with low age probabilities, $A_{k,i}<0.001 \Delta t/\text{yr}$ can be ignored, reducing the computational cost significantly. Our discrete marginalisation of the age parameters implicitly neglects variations in $\mathbf{b}(t)$ on timescales shorter than $\Delta t$. We find, however, that the associated errors (for $\Delta t < 100$ yr) are negligible relative to the overall uncertainty in the data.

\subsubsection{Sedimentary data}
\label{section_SED_data_model}

\begin{figure}[t]
\centering
\includegraphics[width=8.6cm]{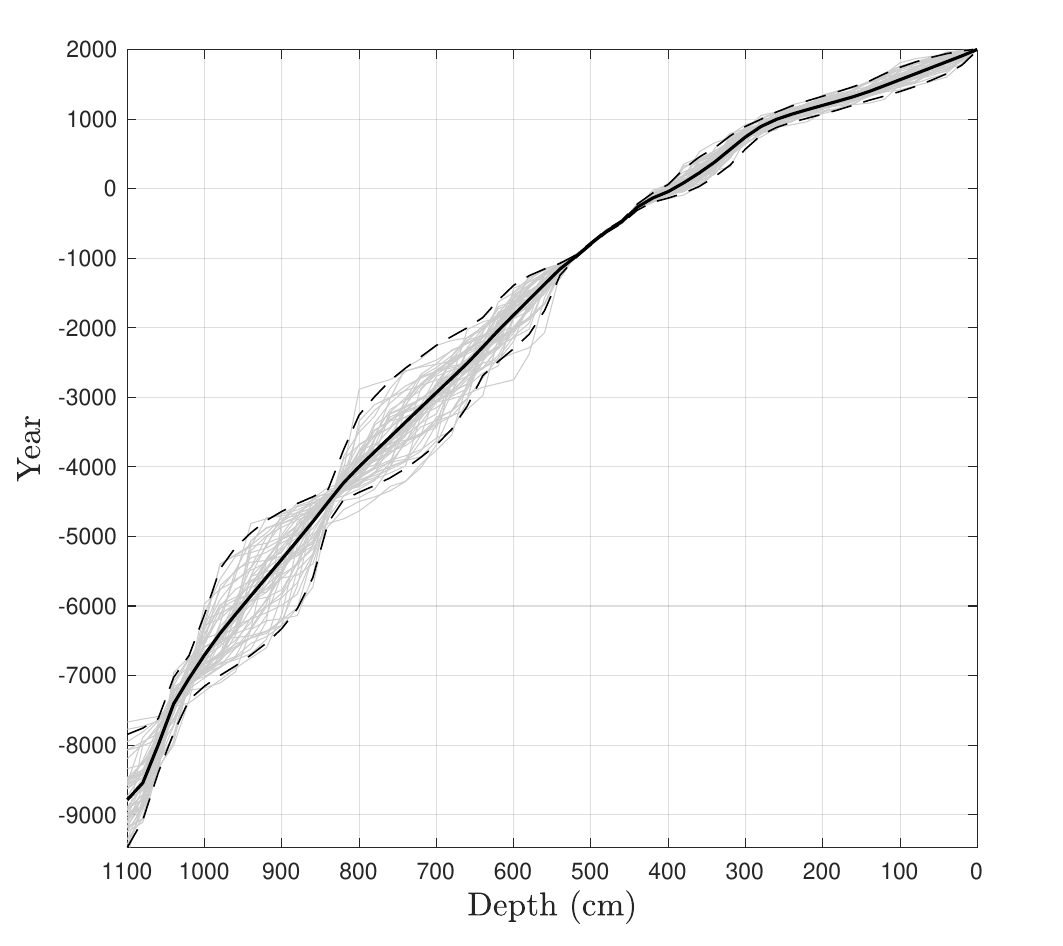}
\caption{The "prior" age-depth model for the Gyltigesjön (GYL) sediment record \citep{Snowball2013,Nilsson2022}. Mean model (solid black line), 95\% credible interval (dashed black lines) and 50 samples drawn from the posterior (gray lines).}
\label{Fig_age_model}
\end{figure}

The likelihood function for the sedimentary palaeomagnetic data differ in a few ways from that used for the archaeomagnetic data. The magnetic field observations (directions only, $D_k,I_k$) are represented by samples collected at different depths in a sediment core. The ages are not independent but are themselves the result of a 'prior' age-depth model (Fig. \ref{Fig_age_model}) that interpolates age estimations at a few horizons using stratigraphic information and prior constraints on the sediment accumulation rate \citep{Blaauw2011,Nilsson2021}. In addition, the recorded magnetic signal is subject to pDRM smoothing and delay and additional systematic uncertainties related to the core orientation \citep[e.g.][]{West2022} as well as inclination flattening \citep[e.g.][]{Blow1978}.

To account for pDRM effects, we use the conceptual model of \textcite{Roberts2004} in which the recorded magnetic field signal $\mathbf{c}(d)$ at a specific depth $(d)$ corresponds to an integration of $\mathbf{b}(t)$ over a period of time after deposition. The process is modelled in the depth domain, where time after deposition at depth $d'$ is expressed as a burial depth $\Delta d=d-d'\ (\ge 0)$. The burial depth $\Delta d$ at which the recording is complete is referred to as the lock-in depth, denoted by $\lambda$. The relative contribution of $\mathbf{b}(t)$ at different burial depths is described by a convolution kernel, $f_{\beta,\lambda}(\Delta d)$, referred to as the lock-in function and given by \citep{Meynadier1996,Nilsson2018}
\begin{align}
    f_{\beta,\lambda}(\Delta d)\propto (\lambda-\Delta d)^{\beta}\qquad (0\le \Delta d\le \lambda),
\end{align}
where $\beta$ is a shape parameter. 
The corresponding cumulative probability function is given by
\begin{equation}
    F_{\beta,\lambda}(\Delta d) = 
    \begin{cases}
        1 - (1-\Delta d/\lambda)^{\beta+1},  & \text{for } 0\le \Delta d\le \lambda\\
        1,  & \text{for } \Delta d> \lambda \, .\\
    \end{cases}
\end{equation}
To reduce correlations to $\beta$, the parameter $\lambda$ is replaced with the half lock-in depth \citep{Hyodo1984}
\begin{equation}
    \gamma = \left(1 - 2^{-\frac{1}{\beta +1}} \right) \lambda,
\end{equation}
which corresponds to the burial depth $\Delta d$ at which 50$\%$ of the pDRM is acquired. 
The site-specific lock-in parameters $\{\gamma_s,\beta_s\}$ are a priori unknown and therefore included as hyperparameters in model, with Gamma prior distributions (table \ref{table_prior}) following \textcite{Nilsson2021}.

In \textcite{Nilsson2021}, the transfer function between the depth domain and the time domain is provided by the co-estimated age-depth model. Since the ages are marginalised in our approach (see below), we use the prior mean age-depth relationship (thick solid line in Fig. \ref{Fig_age_model}) to transfer back and forth between $t_i$ and $d_i$. In practice, a discrete time series of the pDRM filtered magnetic field [$\mathbf{c}_1,...,\mathbf{c}_{N_t}]$ is constructed for each sediment record as a discretised convolution of $\mathbf{b}(d)$ with the kernel $f_{\beta,\lambda}(\Delta d)$ using
\begin{equation}
    \mathbf{c}_j = \sum_{i=j}^{N_t} \Delta F^{\beta,\lambda}_{ji}\, \mathbf{b}_i,
\end{equation}
where $\Delta F_{ij}^{\beta,\lambda}=F_{\beta,\lambda}(d_j-d_{i+1})-F_{\beta,\lambda}(d_j-d_i)$ is the probability mass of the burial depth interval $d_j-d_i<\Delta d<d_j-d_{i+1}$. To ensure that lock-in is always complete, we set $\Delta F^{\beta,\lambda}_{ji} = 1 - F_{\beta,\lambda}(d_j-d_i)$ for all $j$ when $i=N_t$.

Following \textcite{Nilsson2021}, to account for systematic directional errors we introduce the correction parameters $\delta D_s$ and $\delta I_s$ with normally distributed priors (table \ref{table_prior}). These parameters enter eqs. \eqref{chi_sed_dec} and \eqref{chi_sed_inc} below as site-specific corrections of declination and inclination data respectively.

Due to the chronologic co-dependencies of observations sampled at different depths, marginalising their ages is challenging. As more and more measurements from the same core are considered, the number of possible age-depth combinations rapidly becomes intractable. We therefore approach this problem by treating the ages as independent and penalize data with strongly correlated errors introduced by the chronologic uncertainties. In practice this is achieved by inflating the declination and inclination uncertainties by introducing the additional terms $\sigma_{DA}$ and $\sigma_{IA}$, see Appendix \ref{appendix_age_errors}. By considering the ages as independent we can use the same approach used for the archaeomagnetic data, which dramatically reduces the computational cost. The discrete (non-Gaussian) age probabilities, $A_{k,i}$, are obtained empirically from the prior age-depth model samples (Fig. \ref{Fig_age_model}). The normalised model-data residuals, calculated for each model time step $t_i$ using the equivalent notation to eq. \eqref{chi_D} and \eqref{chi_I}, becomes
\begin{align}
    \label{chi_sed_dec}
    & \chi_{k,i}^{(1)} = \frac{D_k - \textsf{H}_{s,D}(\mathbf{c}_i) + \delta D_s}{\sqrt{\sigma_{D,k}^2 + \sigma_{DA,k}^2}},
    \\
    \label{chi_sed_inc}
    & \chi_{k,i}^{(2)} = \frac{I_k - \textsf{H}_{s,I}(\mathbf{c}_i) + \delta I_s}{\sqrt{\sigma_{I,k}^2 + \sigma_{IA,k}^2}}.
\end{align}
The likelihood for the sediment data, ignoring a multiplicative constant, is then written
\begin{equation}
    \label{SED_likelihood}
    \begin{aligned}
    P(\mathbf{o}_k\, |\, \mathbf{b}_1,...,\mathbf{b}_{N_t},\mathbf{A},\gamma_s,\beta_s,\delta D_s,\delta I_s) \propto \\
    \sum_{i=1}^{N_t} A_{k,i} \prod_{j=1}^2
    \exp\left(-{\frac{\left[\chi_{k,i}^{(j)} \right]^2}{2}} \right).
    \end{aligned}
\end{equation}
In contrast to the archaeomagnetic data there is no need to introduce a Student's t error distribution for the likelihood, as the main sources of systematic errors are already accounted for \citep{Nilsson2021}.  
\section{Results}

Using the synthetic data from section \ref{section_data}, two CFF models are constructed spanning the time period 7000 BCE to 2000 CE. For the first model, CFF.MP, the Midpath dynamo simulation used to generate the synthetic data is used to define the prior statistics as well. We set $\Delta t=50$ years, the same as for pfm9k.2 \citep{Nilsson2022}, and $N_v=46$ (retaining 95\% variance in the reduced rank flow vector).

To define the prior statistics for the second model, CFF.NT, we use the dynamo case 24 from \textcite{Aubert2025}, previously also used and labelled as case Neutral\_top1 in \textcite{Rogers2025}. This case operates at a similar level of equivalent dimensional convective power as the Coupled Earth \citep{Aubert2013} and Midpath \citep{Aubert2017} models, and also produces a westward-drifting columnar eccentric gyre through angular momentum conservation in the coupled outer core-inner core-mantle system. It does however use homogeneous codensity flux boundary conditions, i.e. isotropic forcing at the inner-core boundary (ICB) and the CMB, and a higher dimensional equivalent value of the electrical conductivity ($\sigma=9 \times 10^5$ S.m$^{-1}$) than those earlier models, corresponding to upward revised estimates of this parameter for Earth’s core \citep[e.g.][]{Pozzo2022}. Because the dimensional rotational period of this model is about 10 years, flow variability on timescales shorter than $\sim$100 years are not considered Earth like \citep[e.g.][]{Gillet2024}. For this model we therefore set $\Delta t=100$ years and $N_v=54$ (to retain 95\% variance in the reduced rank flow vector).

We note that the amplitude of the time-averaged axial dipole field $\langle |g_1^0| \rangle$ (in relation to the rest of the core field) can vary significantly between different dynamo series. In the case of the two dynamos considered here, $\langle |g_1^0| \rangle$ is also considerable larger (Midpath: $58.0 \pm 4.4\, \mu$T, Neutral\_top1: $45.0 \pm 6.7\, \mu$T) than what is expected for Earth's core over the past 9000 years (pfm9k.2: $32.2 \pm 4.1\, \mu$T). To facilitate sampling, in preparation of applying the method to real data, the prior mean for $g_1^0$ (contained within $\mathbf{b}_0$) is replaced with an estimate that is consistent with the data. For the synthetic tests here we simply use the mean $g_1^0$ from Midpath.

For each model, the HMC sampler was run on four chains with 1000 burn-in iterations, which were discarded, and 500 sampling iterations leading to a total sample size of $M=2000$. The estimates of the effective sample size are larger than 2000 for the majority of the model quantities and the $\hat{R}$ diagnostic statistic \citep{vehtari2021} for all quantities was equal to 1.00, indicating between-chain and within-chain convergence.

\subsection{Model-data comparison}

\begin{figure}[t]
\centering
\includegraphics[width=8.6cm]{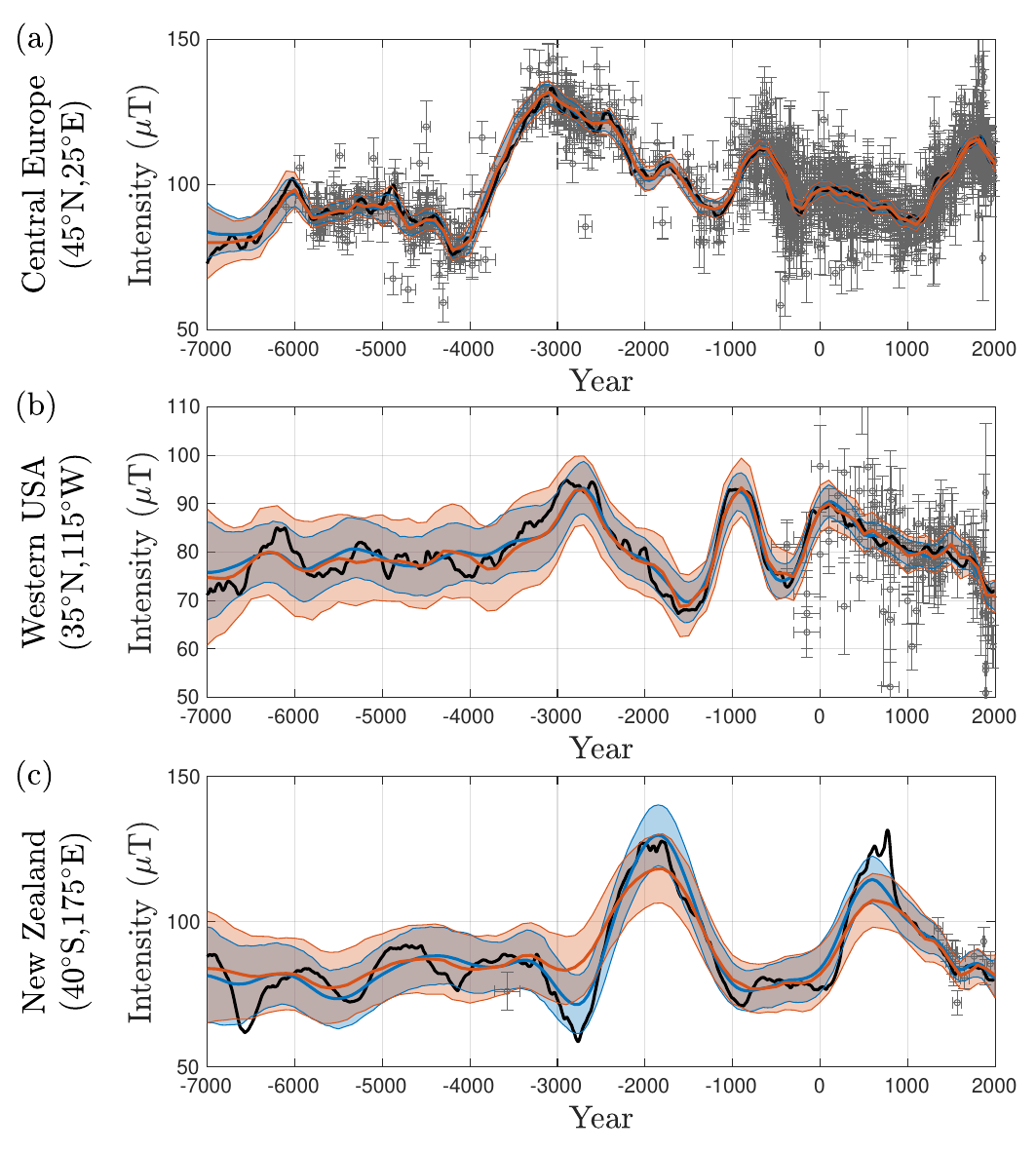}
\caption{Model-data comparison of intensity variations in (a), Central Europe (b). Western USA and (c) New Zealand. The posterior mean (solid lines) and 95\% range (shaded areas) of models CFF.MP based on Midpath prior (blue) and CFF.NT based on Neutral\_top1 (red) are compared to the reference (black lines). The archaeomagnetic data (black dots) were selected from radii of $10^{\circ}$ and relocated assuming an axial dipole field. Error bars denote 1-sigma uncertainties.}
\label{Fig_ARC_model_pred}
\end{figure}

Despite the noisy character of the (synthetic) archaeomagnetic data, MF variations in declination, inclination and intensity, $D,I,F$, on centennial timescales and longer are generally well recovered, particularly in densely populated regions such as Europe (Fig. \ref{Fig_ARC_model_pred}a). In less densely populated regions, e.g. the Western USA (Fig. \ref{Fig_ARC_model_pred}b), the dispersion of the posterior models increase, but the reference time series is generally within the 95\% range. For remote southern hemisphere locations, such New Zealand (Fig. \ref{Fig_ARC_model_pred}c), we note, however, that large and rapid fluctuations away from the prior mean are not always captured. This is particularly true for the CFF.NT model, likely due to a lack of similar dynamics in the Neutral\_top1 prior dynamo series.

\begin{figure*}[t]
\centering
\includegraphics[width=1.0\textwidth]{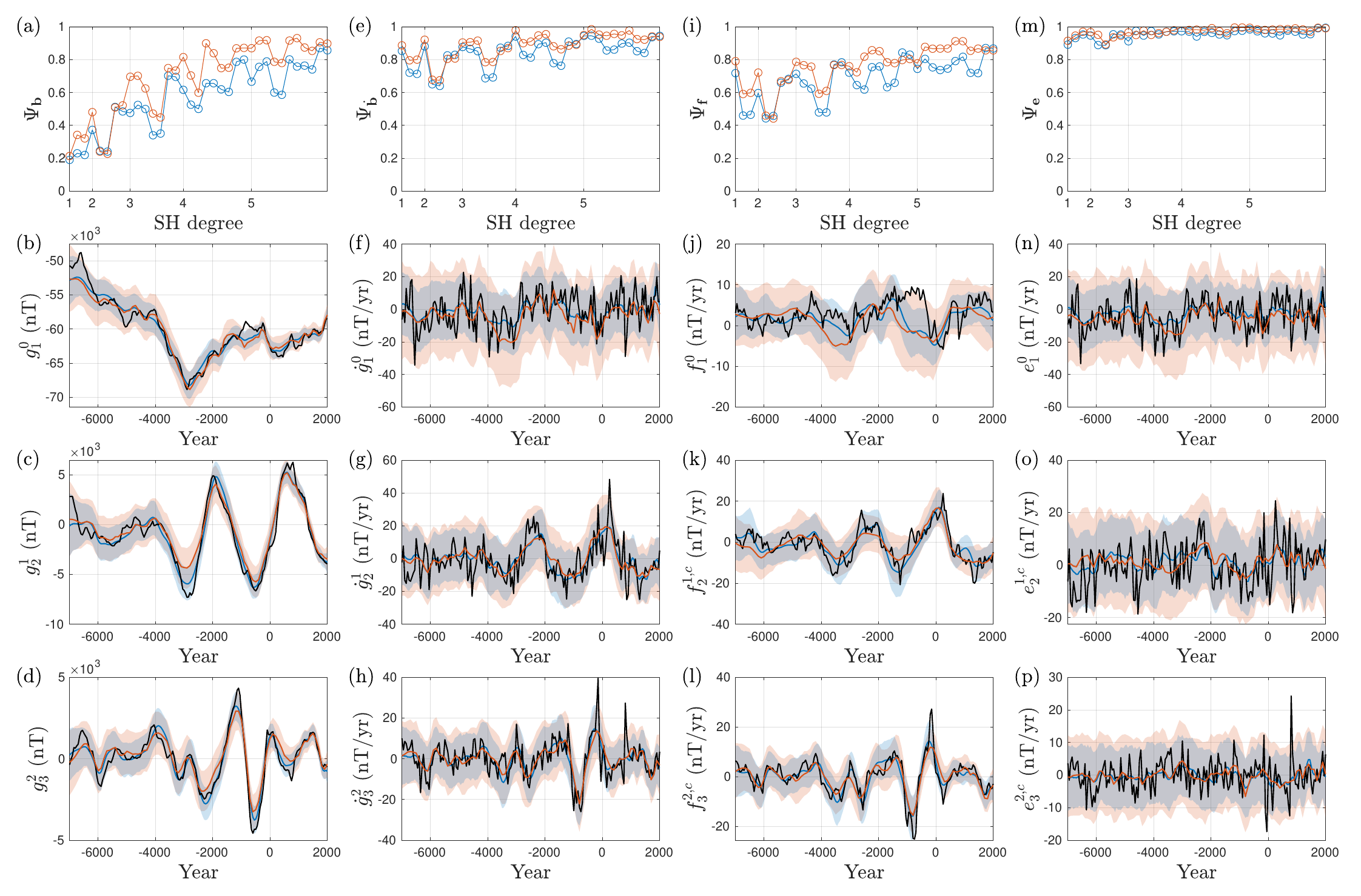}
\caption{Comparison of models CFF.MP based on Midpath prior (blue), CFF.NT based on Neutral\_top1 (red) and the reference time series (black). (a-d) MF $\mathbf{b}$, (e-h) total SV $\dot{\mathbf{b}}$, (i-l) induced SV $\mathbf{f}$ and (m-p) error of representativeness $\mathbf{e}$. The top panel shows the time-integrated posterior-to-prior error ratio $\Psi$ for all coefficients and below we plot time series comparisons of selected coefficients. The CFF models are represented by the posterior mean (solid lines) and 95\% range (shaded areas). For a fairer comparison, the reference time series has been resampled at a 50-yr temporal resolution.}
\label{Fig_MF_SV_coeffs}
\end{figure*}

To simulate the presence of outliers, the synthetic archaeomagnetic data were corrupted Student's t-distributed noise, generated using a reference $\nu=4$ degrees of freedom (\ref{section_data}). The posterior means of $[\nu_1,\nu_2,\nu_3]$, i.e. the modelled number of degrees of freedom of the Student's t error distribution for $D,I,F$ respectively, all fall within a narrow range 5.2 to 5.8 (for both models). These values are slightly high compared to the reference $\nu=4$, but the differences in the resulting error distributions due to this slight overestimation are negligible. The results rather demonstrate (in this idealised example) that the models are able to identify outliers in the data using prior information from the dynamo series.

Due to the marginalisation of the age-depth models and the multiple model parameters involved, model predictions of sedimentary data are less straight forward. As demonstrated by the three examples shown in \ref{appendix_sed_data}, millennial-scale MF variations are generally well recovered by the models. The recovery of shorter term centennial-scale variations is limited due to the often relatively large age uncertainties, but we note that this higher frequency variability is mostly included in the posterior uncertainty envelope.

As previously noted \citep{Nilsson2021,Bohsung2023}, models of pDRM lock-in are generally not sensitive to the shape of the of the lock-in function and in most cases the posterior distribution (of the parameter $\beta$ in our model) is not notably different from the prior distribution (e.g. Fig. \ref{Fig_GYL_model_pred}d). The half lock-in depth, $\gamma$, is more constrained by the data, but for a few records the model significantly under-/overestimate the reference value (see Fig. \ref{Fig_CHM_model_pred}e). This is due to an ambiguity between lock-in delay, resulting in a recorded signal that is younger than the age of the sediments carrying it, and age estimates that are too young compared to the true age, which have a similar apparent effect (for example at 800 cm in Fig. \ref{Fig_CHM_model_pred}a). This ambiguity leads to an underestimation of the half lock-in depth and consequently a slight under-correction of the lock-in smoothing in the data. We note that similar issues also arise in examples with age estimates that are too old, which in turn results in an overestimation of the the lock-in depth (see Fig. \ref{Fig_BIW_model_pred}e) and over-correction of the lock-in smoothing. However, in practice, these effects are small in comparison to other (mostly chronology-related) errors. A similar effect was discussed by \textcite{Nilsson2021}, but we find that the problem is exacerbated here due to the discrete marginalisation of the ages. Systematic errors in declination, $\delta D$, are well recovered by the model and the same is mostly also true for systematic errors in inclination, $\delta I$. However, in a few cases the model significantly under-/overestimate the reference $\delta I$, but as with the lock-in depth, the effects are generally small compared to other sources of errors. 

\begin{figure*}[t]
\centering
\includegraphics[width=1.0\textwidth]{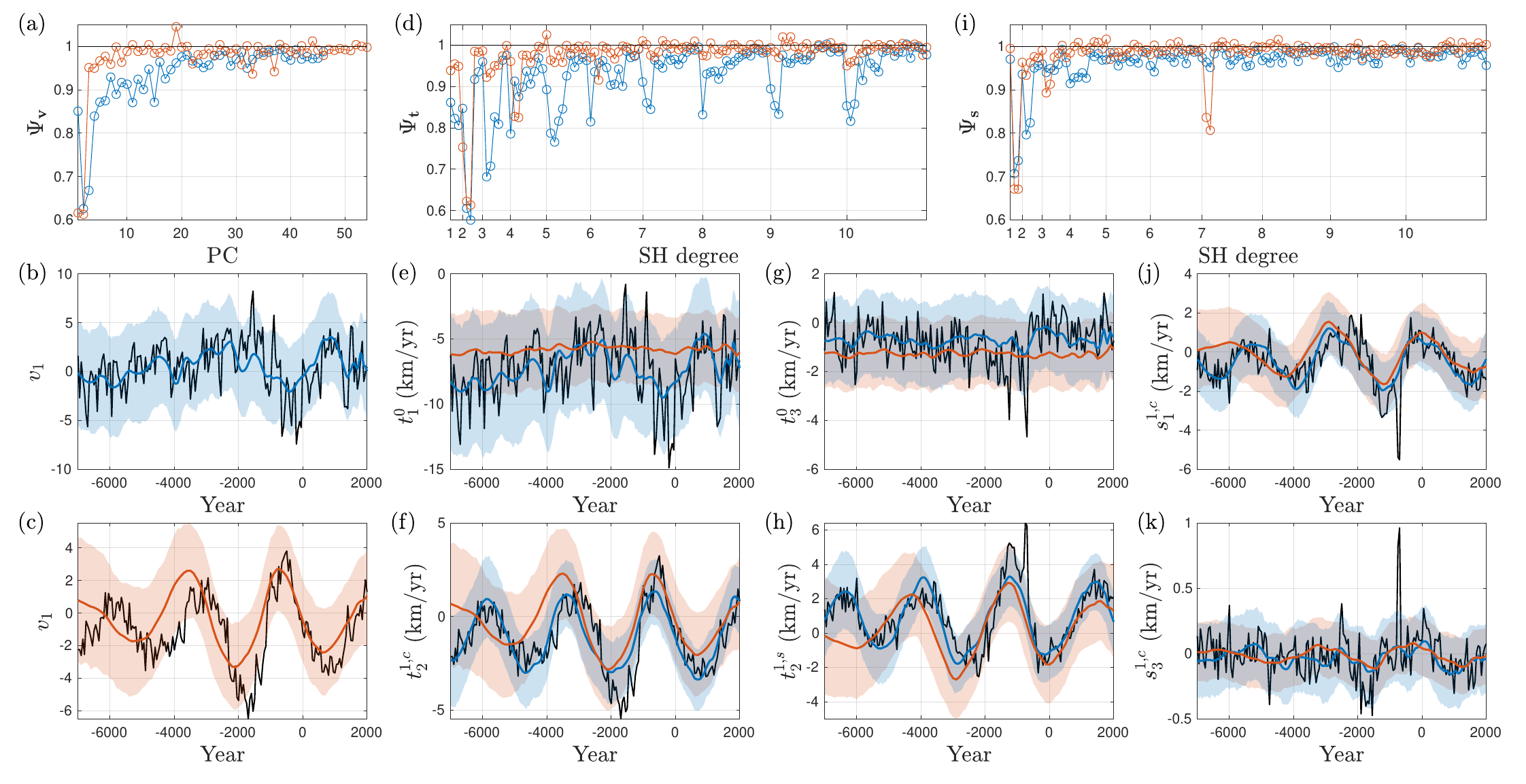}
\caption{Comparison of models CFF.MP based on Midpath prior (blue), CFF.NT based on Neutral\_top1 (red) and the reference time series (black). From left to right: (a-c) Reduced rank flow $\mathbf{v}$, (d-h) toroidal flow $\mathbf{t}$ and (i-k) poloidal flow $\mathbf{s}$. the Top panel shows the time-integrated posterior-to-prior error ratio $\Psi$ for all coefficients and below we plot time series comparisons of selected coefficients. Since the PCA basis depends on the prior, we plot the first component $\mathbf{v}_1$ for each CFF model with their respective reduced basis reference separately. The CFF models are represented by the posterior mean (solid lines) and 95\% range (shaded areas). For a fairer comparison, the reference time series has been resampled at a 50-yr temporal resolution.}\label{Fig_U_coeffs}
\end{figure*}

\subsection{Core field and core flow}

To investigate the models ability to recover core field and core surface flow variations we define $\Psi$ as the time-averaged posterior-to-prior root-mean-square error ratio, for example
\begin{equation}
    \mathbf{\Psi}_{\mathbf{b}} = \frac{1}{N_t}\sum_{i=1}^{N_t} \frac{\sqrt{\Big<(\mathbf{b}_i^{(post)}-\hat{\mathbf{b}}_i)\circ (\mathbf{b}_i^{(post)}-\hat{\mathbf{b}}_i)\Big>}} {\sqrt{\Big<(\mathbf{b}_i^{(prior)}-\hat{\mathbf{b}}_i)\circ(\mathbf{b}_i^{(prior)}-\hat{\mathbf{b}}_i)\Big>}}. 
\end{equation}
Here, $\hat{\mathbf{b}}_i$ represents the reference MF at $t_i$, $\circ$ is the Hadamard product and the angle brackets $\langle\cdot\rangle$ denote ensemble means over 1000 samples drawn from the posterior ${\mathbf{b}}_i^{(post)}$ and the prior $\mathbf{b}_i^{(prior)}$ (generated as in Fig. \ref{Fig_forecast}). The $\Psi$ diagnostic reflects the relative reduction in error achieved by the posterior, with $\Psi=1$ indicating no improvement and $\Psi=0$ complete convergence to the reference.

As demonstrated in Fig. \ref{Fig_MF_SV_coeffs}a-d, the large-scale MF ($l\le3$) is well constrained by both models, but we see only moderate improvements compared to the prior at $l=5$. This is similar to the observations by \textcite{Nilsson2021} and justifies the choice of the MF truncation at $L_b=5$. The CFF.MP model performs better than CFF.NT, but the differences are generally small. The SV (Fig. \ref{Fig_MF_SV_coeffs}e-h) is only indirectly constrained by data and as expected we observe generally moderate improvements to the prior apart from a few large-scale SV coefficients, most notably $\{\dot{g}_1^1, \dot{g}_2^1,\dot{g}_3^2\}$ (as well as the $\dot{h}_l^m$ counterparts), that are comparatively well recovered. This can be traced to the induced part of the SV (Fig. \ref{Fig_MF_SV_coeffs}i-l) which contributes strongly to these particular coefficients. The opposite is true for the axisymmetric SV, e.g. $\dot{g}_1^0$ and $\dot{g}_3^0$, where the induced part is small and generally not well recovered by the models. For these coefficients the error of representativeness (Fig. \ref{Fig_MF_SV_coeffs}m-p), which includes sub-grid processes and diffusion, plays a more important role, as previously noted by \textcite{Gillet2019}.

Turning to the core surface flow (Fig. \ref{Fig_U_coeffs}), the CFF.MP model generally outperforms the CFF.NT model. This is not surprising given that the reference was generated using the Midpath geodynamo series. The difference in model performance is exacerbated by the partial decoupling between the core surface flow and the archaeo/palaeomagnetic MF observations \eqref{b_i+1}. This leads to a greater the reliance on the prior statistics being compatible with the true (in this case the reference) core surface flow that we want to recover.

The time-averaged posterior-to-prior error ratio for the reduced rank flow (Fig. \ref{Fig_U_coeffs}a-c) provides a good measure on how much the archaeo/palaeomagnetic data actually constrain the core surface flow. For the CFF.MP model we observe (mostly small) improvements for the first $\sim$20 principal components, compared to the first $\sim$10 for CFF.NT. Note that the PCA basis functions are not the same for CFF.MP and CFF.NT. In both cases the first three components, which explain 57\% and 54\% of the core surface flow variances, respectively, are linked to the large-scale expressions of an eccentric planetary gyre that dominates the flow dynamics in both dynamo simulations (see section \ref{section_gyre} for more discussion on this).

Overall, the archaeo-/palaeomagnetic data are primarily providing constraints on the toroidal flow coefficients $t_l^m$ of orders $m=0$ and $m=1$ (Fig. \ref{Fig_U_coeffs}d-h). As shown in Fig. \ref{Fig_U_coeffs}, the first principal component $v_1$ for CFF.MP is dominated by $t_1^0$ (Fig. \ref{Fig_U_coeffs}b,e), whilst $v_3$ and $v_2$ mainly correspond to $t_2^{1,c}$ and $t_2^{1,s}$ (Fig. \ref{Fig_U_coeffs}f,h). The same is true for CFF.NT, but due to the different PCA basis the corresponding order of the principal components is $\{v_3,v_1,v_2\}$. The $\{t_2^{1,c},t_2^{1,s}\}$ flow coefficients are well constrained by both models, whilst $t_1^0$ is relatively poorly resolved by CFF.NT. This is due to the statistical prior being incompatible with the reference $t_1^0$ time series in terms of variance and correlation time. The poloidal flow coefficients (Fig. \ref{Fig_U_coeffs}i-k) are generally less well constrained. Notable exceptions are the $s_1^{1,c}$ and $s_1^{1,s}$ that are linked to large scale equatorial upwelling and downwelling around eastern and western limbs of the planetary-scale gyre.

\subsection{Weak-field anomalies and eccentric gyres}
\label{section_gyre}

\begin{figure*}[t]
\centering
\includegraphics[width=1.0\textwidth]{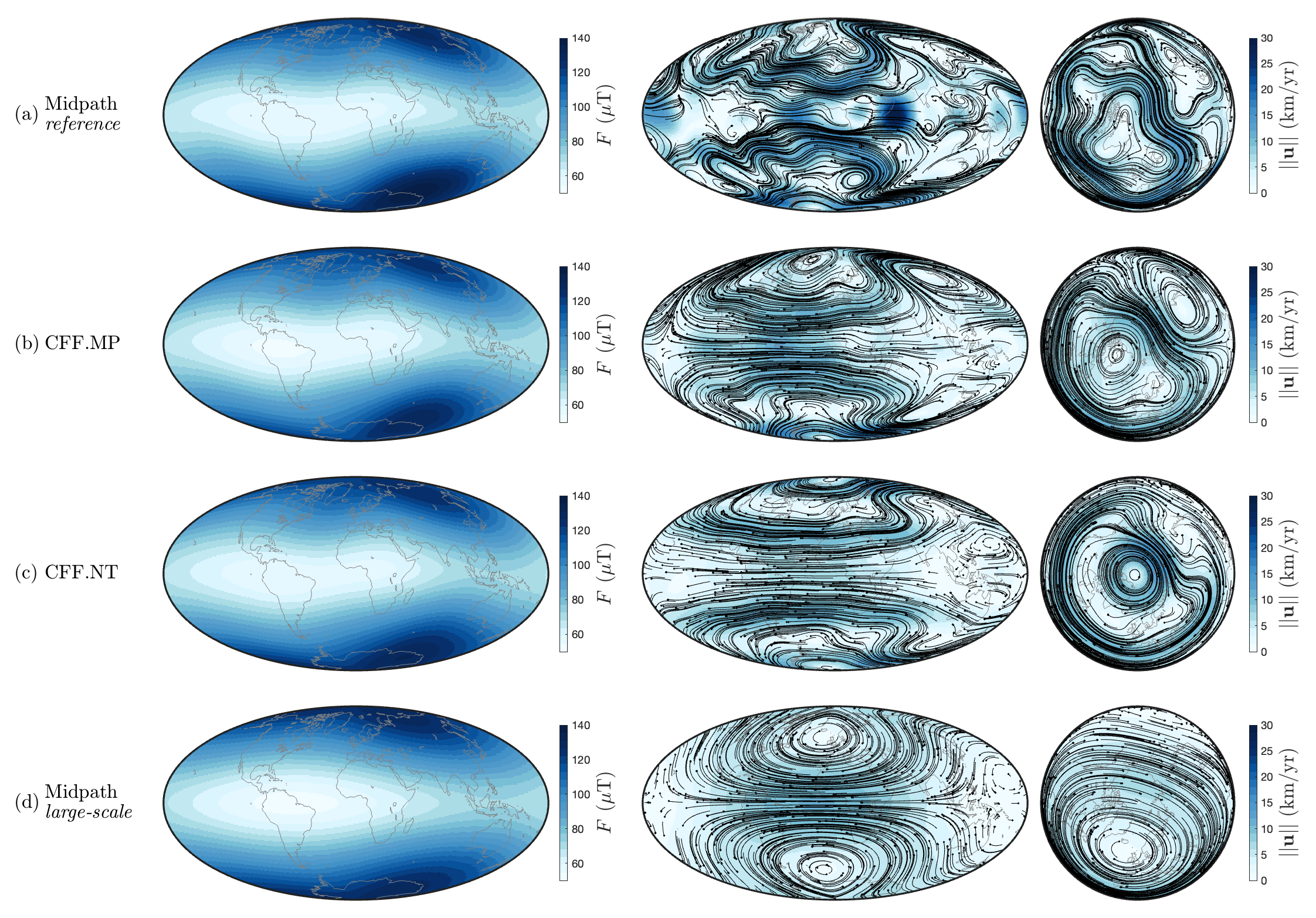}
\caption{Maps of geomagnetic field intensity at Earth’s surface and core surface flow at 1000 CE based on the reference geodynamo series. From top to bottom: (a) the Midpath reference dynamo series, (b) CFF.MP mean model, (c) CFF.NT mean model and (d) large-scale Midpath reference based exclusively on the MF coefficients $\{g_1^0, g_2^1, h_2^1\}$  and toroidal flow coefficients $\{t_1^0, t_2^{1,c}, t_2^{1,s}\}$. The core surface flow direction, indicated by the particle flow head (black dots), is dominantly westward in all maps.}
\label{Fig_gyres_anomalies}
\end{figure*}

In Fig. \ref{Fig_gyres_anomalies}a we show geomagnetic field intensity at Earth's surface and the core surface flow at 1000 CE, according to the reference geodynamo series. The field is dominated by a weak field anomaly with its centre around South America, reminiscent of the present-day South Atlantic Anomaly. The core surface flow is characterized by a planetary-scale axially symmetric eccentric gyre, similar to what is observed in modern core flow inversions \citep[e.g.][]{Gillet2019}, which has its centre around the same longitudes as the weak field anomaly. As demonstrated in Fig. \ref{Fig_gyres_anomalies}b-c, the CFF models are able to recover the large-scale morphology of both the weak field anomaly at Earth's surface and eccentric core surface flow gyre. The CFF.MP model performs slightly better, but overall the level of detail in the two model predictions is comparable and the example illustrates what we can expect to recover based on archaeo/palaeomagnetic data.

\begin{figure*}[t]
\centering
\includegraphics[width=1.0\textwidth]{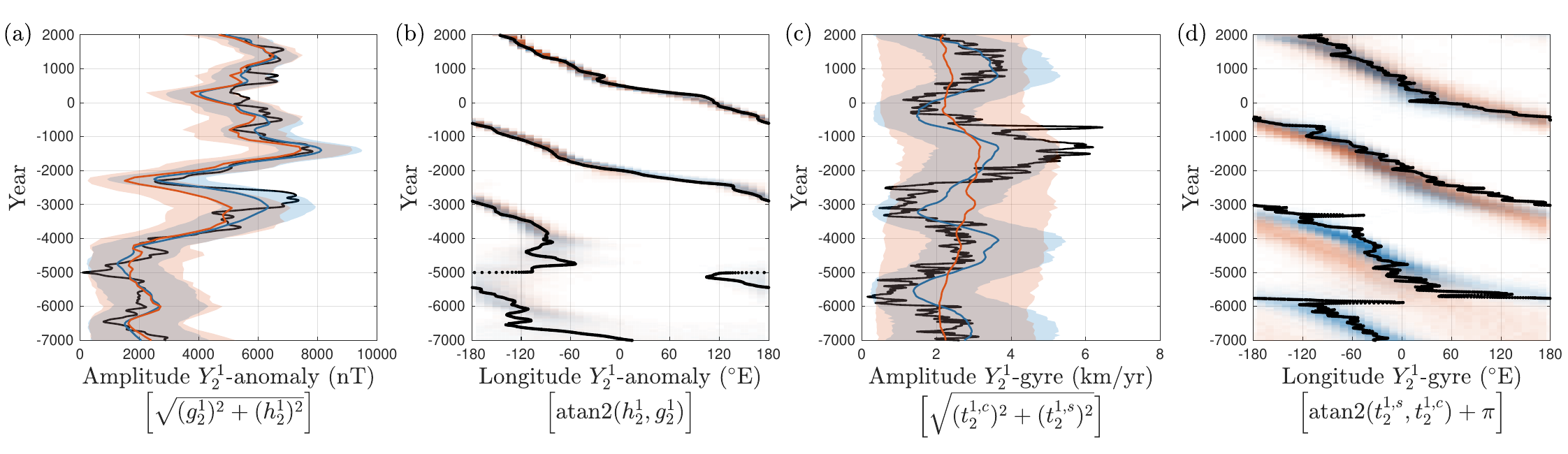}
\caption{Time series and time-longitude density plots of (a) $Y_2^1$-anomaly amplitude, (b) $Y_2^1$-anomaly centre location, (c) $Y_2^1$-gyre amplitude and (d) $Y_2^1$-gyre centre location: CFF.MP (blue), CFF.NT (red) and the reference time series (black). The CFF models in (a,c) are represented  by the posterior mean (solid lines) and 95\% range (shaded areas). For a fairer comparison, the reference time series has been smoothed with a 50-yr running average.}
\label{Fig_TL_gyres_anomalies}
\end{figure*}

It is not a coincidence that the reference dynamo series at 1000 CE shows a weak field anomaly at Earth's surface coinciding with eccentric gyre at roughly the same longitudes. Both the weak field anomaly and the eccentric gyre are semi-persistent features of the Midpath dynamo series that continuously drift westward and appear to be phase-locked to each other. Studying how these type of features evolve over time will be an important application of the CFF models when applied to real data. To track the weak field anomaly, we follow \textcite{Nilsson2022} and introduce a large scale field representation based exclusively on coefficients $\{g_1^0, g_2^1, h_2^1\}$ (Fig. \ref{Fig_gyres_anomalies}d). Assuming a negative $g_1^0$, the amplitude (or size) of the resulting "$Y_2^1$-anomaly" (East-West field asymmetry) is defined as $\sqrt{(g_2^1)^2+(h_2^1)^2}$ and the longitude of its centre as $\atantwo (h_2^1,g_2^1)$. In a similar way we use a large scale representation of the core surface flow, based on the toroidal flow coefficients $\{t_1^0, t_2^{1,c}, t_2^{1,s}\}$ (Fig. \ref{Fig_gyres_anomalies}d), to characterise the evolution of eccentric planetary gyre. Assuming a negative $t_1^0$, the amplitude (or intensity) the resulting "$Y_2^1$-gyre" is defined as $\sqrt{(t_2^{1,c})^2+(t_2^{1,c})^2}$ and the longitude of the centre as $\atantwo (t_2^{1,s},t_2^{1,c})+\pi$.

In Fig. \ref{Fig_TL_gyres_anomalies} we plot variations in amplitude and location of the reference $Y_2^1$-anomaly and $Y_2^1$-gyre and compare these to CFF model predictions. Apart from a slight underestimation of the amplitude around 3000 BCE, both models are successfully able to track changes in the $Y_2^1$-anomaly over the past 9000 years. The westward drift of the anomaly, manifesting as diagonal lines from the bottom right to top left (Fig. \ref{Fig_TL_gyres_anomalies}b), is briefly interrupted around 5000 BCE when the anomaly itself briefly disappears. Variations in the $Y_2^1$-gyre are unsurprisingly more dependent on the prior. Both models are able to track the gyre’s location over the time interval 2000 BCE - 2000 CE, which contains 86\% of the data. For the earlier less well constrained epochs, the CFF.NT model prediction deviates slightly from the reference, reverting to a the steady-state westward drift of $\sim0.09^{\circ}$/yr of the Neutral\_top1 prior. The prior dependency is most noticeable for the amplitude of the $Y_2^1$-gyre, with the CFF.NT model appearing mainly sensitive to variations on multi-millennial timescales. This is related to a dynamic interplay between the $\{t_2^{1,c},t_2^{1,s}\}$ and $t_1^0$ coefficients in the Midpath reference geodynamo series which is not captured by the CFF.NT model due to the long correlation time of the $t_1^0$ in the Neutral\_top1.
\section{Discussion}

Here we present the first ever integrated core field and core flow modelling framework for archaeo/palaeomagnetic MF observations and test it using synthetic data. The results demonstrate the possibility to retrieve information on large scale core surface flow structures such as the planetary scale gyre observed in modern core flow inversions \citep[e.g.][]{Pais2008,Gillet2019,Aubert2014}. The reduced basis flow, as opposed to truncating the core flow expansion, allows the model to capture Earth-like core flow structures with a limited number of parameters as illustrated in Fig. \ref{Fig_gyres_anomalies}. For the two prior dynamo series used here, the first three principal components are dominated by toroidal flow coefficients $\{t_1^0, t_2^{1,c}, t_2^{1,s}\}$, but it is evident from the results that higher degree and order core flow coefficients provide important contributions.

The tests highlight the influence of the dynamo prior \citep[see also][]{Suttie2025}. The reliance on kinematic statistics from numerical dynamo series is both a strength and a weakness of our method. Reconstructing millennial scale variations in core flow based on sparse and noisy archaeo/palaeomagnetic observations is an inherently non-unique problem \citep{Holme2015}. Such reconstruction will inevitably involve assumptions on (i) the role of diffusion, (ii) the influence of unmodelled sub-grid processes and (iii) the temporal and spatial characteristics of the core surface flow. Physical constraints such as quasi-geostrophy \citep{Pais2008} can help reduce the ambiguity of core-surface flow inversions, but they do not eliminate it entirely, even for spatially well determined magnetic field \citep{Scwaiger2023}. Given the severity of this non-uniqueness for core flow reconstructions on millennial timescales, we argue that reliance on some additional information from numerical dynamo simulations is unavoidable. In this context, using numerical geodynamo simulations to define the prior statistics represents a set of physically motivated but rather strong assumptions about the dynamics of Earth's core.

The recovered flow will be influenced, and to some degree limited, by the choice of the prior geodynamo series. This is clearly demonstrated by the CFF.NT models ability to capture variations in the amplitude of the planetary gyre (Fig. \ref{Fig_TL_gyres_anomalies}c). In the (reference) Midpath geodynamo series, the planetary gyre oscillates in and out of eccentric phases, growing in intensity as the eastern equatorward limb passes over a zone of preferential upwelling beneath Indonesia, due to differential growth of the inner core \citep{Aubert2013}. In contrast, The Neutral\_top1 dynamo is run with isotropic boundary forcing which leads to a steadily westward drifting gyre that exhibits much less variation through time and hence the CFF.NT model also fails to capture these type of variations. On the other hand, as demonstrated by \textcite{Suttie2025}, using a Coupled Earth dynamo series such as Midpath, to define the prior statistics may bias the zonal structure of the recovered flow, e.g. with eccentric gyres forming preferentially in the Atlantic Hemisphere. To address this issue, we suggest to construct several different CFF models using an ensemble of different geodynamo series, with different boundary conditions, to define the prior statistics. A comparison of the results from such a survey will provide a better constraint on what features are data-driven and not \citep{Rogers2025}.

By marginalising as opposed to co-estimating \citep{Nilsson2021} the archaeomagnetic ages, we dramatically reduce the number of model parameters and avoid convergence issues for the HMC sampler related to multimodal posterior age distributions \citep[see Fig. 7,][]{Nilsson2021}. Independent comparisons of the two methods (reproducing pfm9k.2) show that there are only negligible differences in the recovered geomagnetic field. The main reason for this is that the age uncertainties of the archaeomagnetic data included in the model are in most cases small enough that the added geomagnetic field information does not provide much additional chronologic constraints. This is evident from the high (0.91) root mean square normalised dispersion, defined as the standard deviation of the posterior divided by the standard deviation of the prior, obtained by \textcite{Nilsson2021} for the archaeomagnetic age parameters. In other words, the co-estimated (prior) age distributions are in most cases not updated by the geomagnetic field data and therefore the results are essentially equivalent to integrating over the priors, i.e. marginalising the ages. Our method performs similar to transferring the age uncertainties into observation uncertainties \citep{Schanner2021,McHutchon2011}, but has the advantage of not requiring normally distributed age errors. In future applications, the method can easily be adapted for other age-error distributions that are consistent with the archaeological constraints \citep[e.g.][]{Brown2021}.

For the sedimentary data, the marginalisation of the ages is also improving the HMC sampling by avoiding convergence issues \citep[see Fig. S6,][]{Nilsson2022}. However, this comes with a loss of information due to the treatment of the ages as independent, i.e. ignoring stratigraphy. To prevent overfitting data with strongly correlated age errors the declination and inclination uncertainties are inflated (see \ref{appendix_age_errors}). As a results, the models are less sensitive to sediment parameters related to pDRM lock-in and systematic declination/inclination errors. In future applications these parameters could potentially be replaced with fixed quantities, obtained using a pre-processing procedure constrained by archaeomagnetic data \citep{Bohsung2024}. This would further alleviate computations and may also avoid introducing systematic errors due to over/underestimations of pDRM lock-in. The main benefit of the new approach of treating chronological uncertainties is that it will enable adding more sediment records to the model.

\begin{acknowledgements}
The research was funded by the European Union (ERC, PALEOCORE, 101125394). Views and opinions expressed are however those of the authors only and do not necessarily reflect those of the European Union or the European Research Council. Neither the European Union nor the granting authority can be held responsible for them. AN and NS also acknowledge funding from the Swedish Research Council (DNR2020–04813). NG and JA acknowledge funding by ESA in the framework of EO Science for Society, through contract 4000148713/25/NL/FFi (4D Earth Core+). This project was financially supported by CNES as an application of the Swarm mission. NG is part of Labex OSUG@2020 (ANR10 LABX56).
\end{acknowledgements}

\section*{Data availability}
The code and input data files needed to run the CFF models as well as the output files can be found at
\break
 \href{https://earthref.org/ERDA/2776/}{https://earthref.org/ERDA/2776/}.

\section*{Competing interests}  
The authors have no competing interests.

\printbibliography

\appendix
\counterwithin{figure}{section}
\appendixpage
\section{Correlated age errors}
\label{appendix_age_errors}

The additional uncertainty terms $\sigma_{DA,k}$ and $\sigma_{IA,k}$ (section \ref{section_SED_data_model}) are introduced to account for correlated observation errors as a result of correlated age errors. We note that while the age errors for a given sediment record are generally well approximated by a multivariate normal distribution, the resulting observation errors are not and therefore cannot be treated with conventional methods. A consequence of the stratigraphic constraint (the monotonicity of the age model) is that large age errors are inevitably also highly correlated. The magnitude of the correlation in the resulting observation error depends additionally on the rate-of-change of the recorded geomagnetic field. Age errors that are small compared to the timescale of field variability will result in highly correlated but generally small observation errors that are unproblematic. Problems arise when the age errors are of similar magnitude or larger than the timescale of the recorded field variations. Motivated by these observations, we define $\sigma_{DA,k}$ and $\sigma_{IA,k}$ as the age uncertainty weighted by a prior estimate of the SV for the given field component at the site coordinates \citep[see eq. 24,][]{Schanner2021}
\begin{align}
    & \sigma_{DA,k} = \sigma_{a,k} \, \sigma_{p\dot{D},k},
    \\
    & \sigma_{IA,k} = \sigma_{a,k} \, \sigma_{p\dot{I},k},
\end{align}
where $\sigma_{a,k}^2$ is the variance of the prior age model at the depth of the sample and $\sigma_{p\dot{D},k}^2$ and $\sigma_{p\dot{I},k}^2$ are the prior estimates of the SV variance of declination and inclination respectively. We note that this definition of the additional uncertainty terms is the same as the transformation of age uncertainties to measurement uncertainties used by \textcite{Schanner2021} to address age errors for archaeomagnetic data.

\begin{figure*}[ht]
\centering
\includegraphics[width=\textwidth]{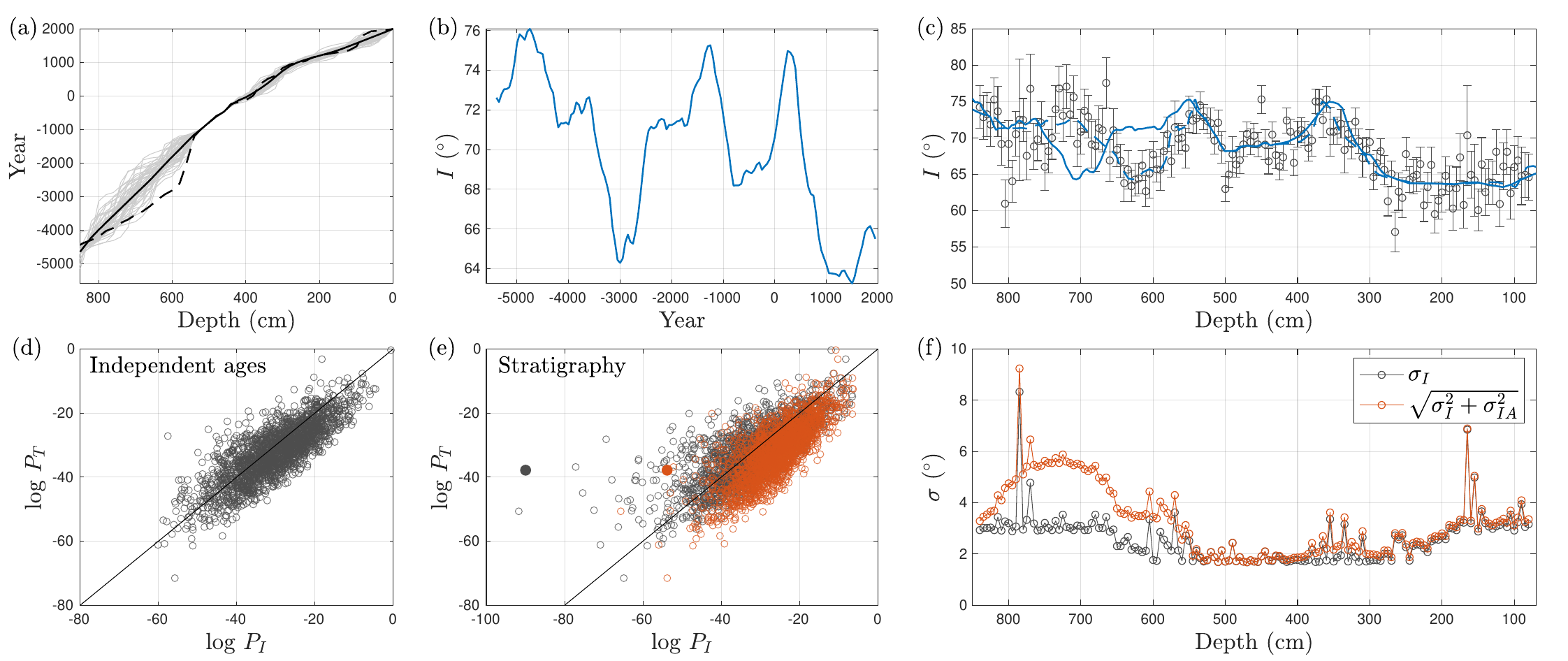}
\caption{Tests illustrating the observational error introduced by correlated age errors using synthetic data based on the palaeomagnetic record from Gyltigesjön \citep{Snowball2013,Nilsson2022}. The upper panel shows an example case: (a) The reference age-depth model (dashed line) drawn from the prior age-depth model (mean model shown as solid black line and 50 samples as thin gray lines). (b) The reference inclination time series $\hat{I}(t)$ (blue line) drawn from a Gaussian Process with a Matern-$\sfrac{3}{2}$ kernel function (see text for more details). (c) Synthetic inclination data generated using the reference timeseries and reference age-depth model and additionally corrupted with Gaussian noise scaled by $\sigma_I$. The reference time series is projected onto the data depth-scale based on both the mean age-depth model (solid blue line) as well as the "true" reference age-depth model (dashed blue line). The lower panel: Scatter plot of log-probabilities $\log P_T$ and $\log P_I$ for 2000 cases generated as above (d) without and (e) with stratigraphic constraints. For the latter $\log P_I$ is evaluated using both $\sigma_I$ (black circles) and $\sqrt{\sigma_I^2 + \sigma_{IA}^2}$ (red circles) as shown in (f). For illustration purposes, the $P_T$ and $P_I$ probabilities were normalised so that the respective sums of the 2000 cases equals one. The case from the upper panel is shown as solid black/red circles in (e).}
\label{Fig_SED_data}
\end{figure*}

In principle, $\sigma_{p\dot{D},k}^2$ and $\sigma_{p\dot{I},k}^2$ can be obtained empirically from the prior dynamo simulation. However, in cases with pDRM smoothing, this will result in an overestimation of the recorded SV variance. With increasing pDRM smoothing the correlated age uncertainties become less problematic and we therefore estimate the SV variance directly from the data using Gaussian Process regression \citep{Rasmussen2006}. We consider the inclination data from a sediment record (after subtracting the mean and assigning ages based on the prior mean age-model, Fig. \ref{Fig_SED_data}a) as a realization of a zero-mean Gaussian Process with a Matern-$\sfrac{3}{2}$ kernel function \citep{Gillet2013,Allington2023} 
\begin{equation}
    K_{pI}(t,t') = \sigma_{pI}^2 \left(1+\frac{|t-t'|}{\tau_{pI}} \right)\, \exp\left(-\frac{|t-t'|}{\tau_{pI}}\right).
\end{equation}
The parameters $\sigma_{pI}$ and $\tau_{pI}$ are obtained (for each sediment record) through a 2D grid search maximizing the log marginal likelihood \citep[see eq. 2.30,][]{Rasmussen2006}, with time series of inclination  from the prior dynamo simulation at the site coordinates providing the upper and lower bounds of $\sigma_{pI}$ and $\tau_{pI}$ respectively. The derivative of the Gaussian Process is another Gaussian Process with the kernel function $\partial t \partial t' K_{pI}(t,t')$, from which the SV variance is obtained as follows
\begin{equation}
    \sigma_{p\dot{I},k}^2 = \frac{\sigma_{pI}^2}{\tau_{pI}^2}.
\end{equation}
To account for the co-dependence to inclination data we use the following approximation of the SV variance of declination data
\begin{equation}
    \sigma_{p\dot{D},k}^2 = \left( \frac{\sigma_{p\dot{I}}} {\cos{I_k}} \right)^2.
\end{equation}

To illustrate the effects of ignoring stratigraphy, i.e. treating the age errors as independent, we use the  palaeomagnetic record from Gyltigesjön \citep{Snowball2013,Nilsson2022} to generate synthetic data (Fig. \ref{Fig_SED_data}). First, we draw 2000 random inclination time series $\hat{I}(t)$ from the Gaussian Process as defined above (Fig. \ref{Fig_SED_data}b). For each time series a random age-depth model (Fig. \ref{Fig_SED_data}a) is used to generate synthetic inclination data at the depths of measurements, corrupted by Gaussian noise according to the measurement uncertainties $\sigma_{I,k}$ (Fig. \ref{Fig_SED_data}c). In parallel, we construct 2000 equivalent datasets where the stratigraphy is ignored by randomly drawing the ages for each measurement from different age-depth models, resulting in truly independent age errors. For each case of $\hat{I}(t)$ we calculate the probability or likelihood (ignoring constant terms) of the $N_{GYL}$ data with known (true) ages:
\begin{equation}
    \log P_{T} = \sum_{k=1}^{N_{GYL}} -\frac{\left( I_k - \hat{I}(t_k) \right)^2}{2\sigma_{I,k}^2},
\end{equation}
and by discretely integrating out the ages, equivalent to \eqref{SED_likelihood}, over $N_t$ time steps (50-yr resolution)
\begin{equation}
    \log P_{I} = \sum_{k=1}^{N_{GYL}} \log \left[ \sum_{i=1}^{N_t} A_{k,i}\, \exp\left(-\frac{\left( I_k - \hat{I}(t_i) \right)^2}{2\sigma_{I,k}^2}\right) \right].
\end{equation}
The scatter plot of $\log P_T$ vs $\log P_I$ for the 2000 cases with independent ages (Fig. \ref{Fig_SED_data}d) shows the expected effect of age errors. Despite fairly large age uncertainties, the log probabilities are on average proportional to each other. However, when considering the stratigraphic data (Fig. \ref{Fig_SED_data}e, black circles), large deviations from proportionality are observed due to the effect of correlated age errors. This is clearly demonstrated by the highlighted example in the upper panel of Fig. \ref{Fig_SED_data}, where correlated age errors between 800 and 550 cm produce apparent systematic observation errors resulting in anomalously low $\log P_{I}$ (filled black circle in Fig. \ref{Fig_SED_data}e). Replacing $\sigma_{I,k}$ with $\sqrt{\sigma_{I,k}^2+\sigma_{IA,k}^2}$ (red circles in Fig. \ref{Fig_SED_data}e-f) leads to a general flattening of $\log P_I$ with the effects primarily concentrated to the problematic parts of the data. In other words, the data with a tendency to have correlated age errors are down-weighted in the likelihood function.
\section{Model predictions of sedimentary data}
\label{appendix_sed_data}

\begin{figure*}[ht]
\centering
\includegraphics[width=1.0\textwidth]{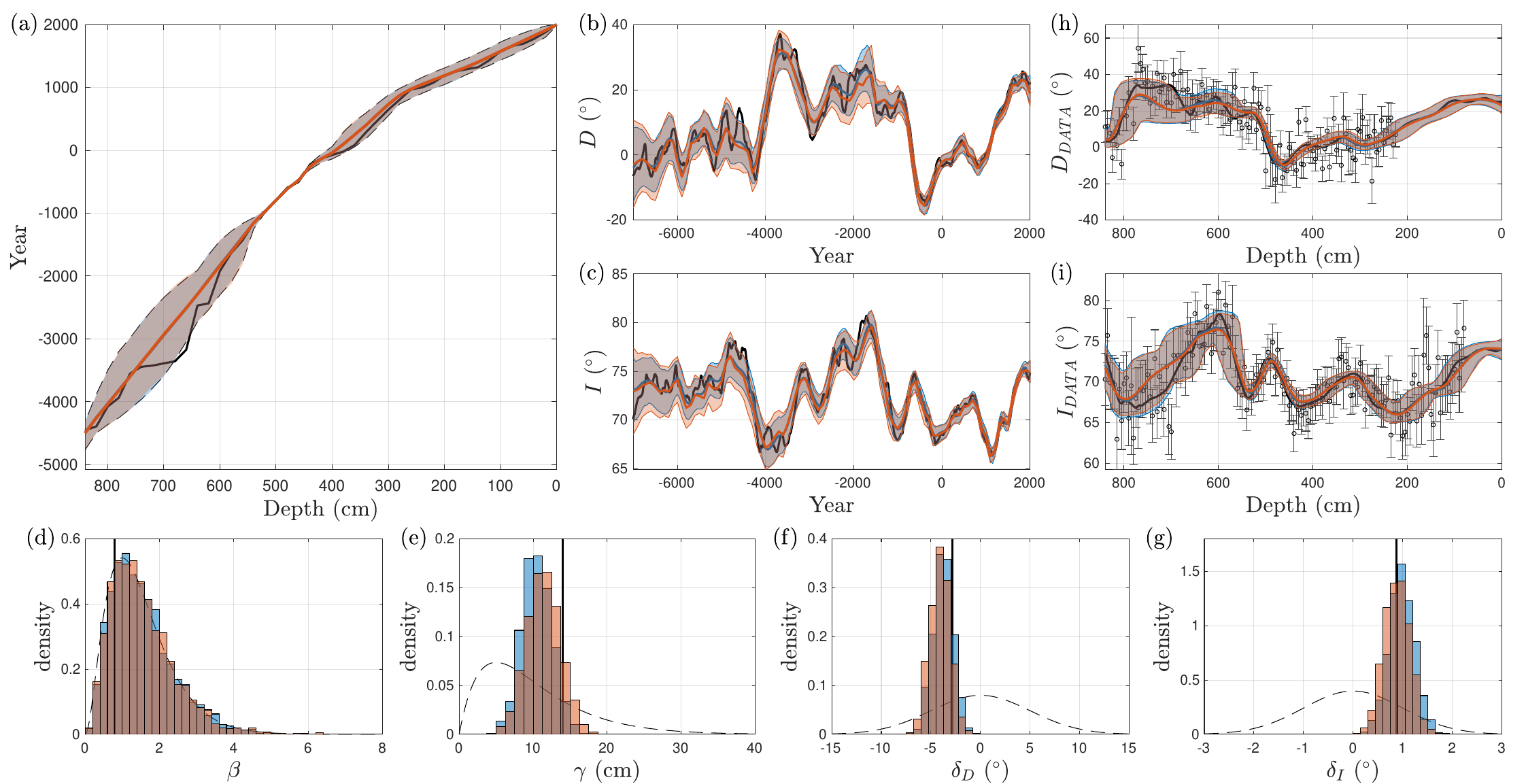}
\caption{Recovery of model parameters associated with the Gyltigesjön (GYL) sediment record. CFF models based on Midpath prior (blue), Neutral\_top1 (red) are compared to the reference (black lines) and the prior distribution (dashed black lines), where appropriate. (a) Age-depth, (b-c) model predictions of $D(t)$ and $I(t)$ for the site coordinates, (d) lock-in shape $\beta$, (e) half lock-in depth $\gamma$, systematic errors in (f) declination $\delta_D$ and (g) inclination $\delta_I$ and (h-i) model-data comparison of $D_k$ and $I_k$. Due to the marginalization of the ages, the model predictions of $D_k$ and $I_k$ are not easy to illustrate. We therefore approximate them here by transferring the predicted $D(t)$ and $I(t)$, including pDRM effects and systematic errors, to depth-scale use random samples drawn from the prior age-depth model.}
\label{Fig_GYL_model_pred}
\end{figure*}

\begin{figure*}[t]
\centering
\includegraphics[width=1.0\textwidth]{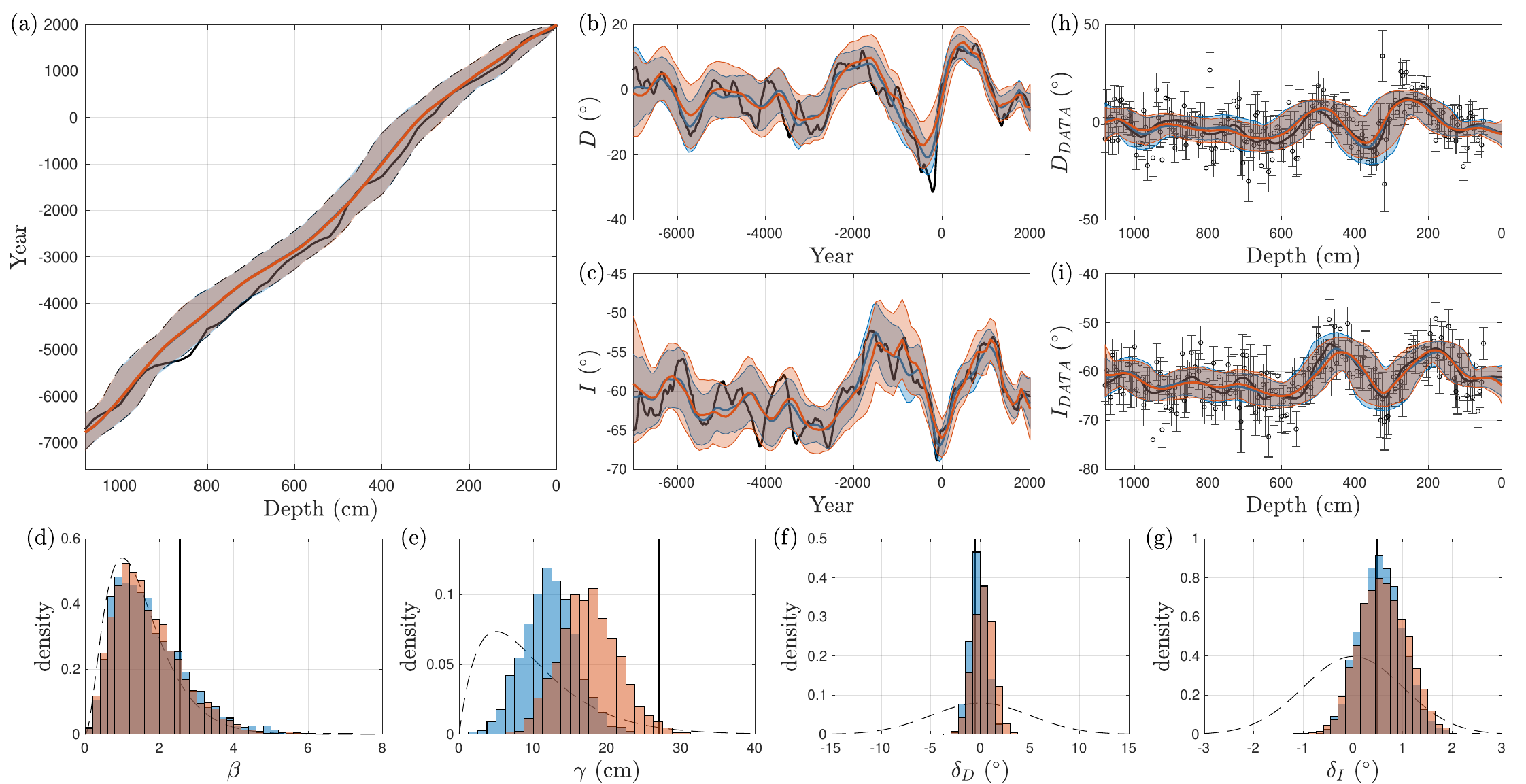}
\caption{Recovery of model parameters associated with the Chilean margin (CHM) sediment record. CFF models based on Midpath prior (blue), Neutral\_top1 (red) are compared to the reference (black lines) and the prior distribution (dashed black lines), where appropriate. (a) Age-depth, (b-c) model predictions of $D(t)$ and $I(t)$ for the site coordinates, (d) lock-in shape $\beta$, (e) half lock-in depth $\gamma$, systematic errors in (f) declination $\delta_D$ and (g) inclination $\delta_I$ and (h-i) model-data comparison of $D_k$ and $I_k$. Due to the marginalization of the ages, the model predictions of $D_k$ and $I_k$ are not easy to illustrate. We therefore approximate them here by transferring the predicted $D(t)$ and $I(t)$, including pDRM effects and systematic errors, to depth-scale use random samples drawn from the prior age-depth model.}
\label{Fig_CHM_model_pred}
\end{figure*}

\begin{figure*}[t]
\centering
\includegraphics[width=1.0\textwidth]{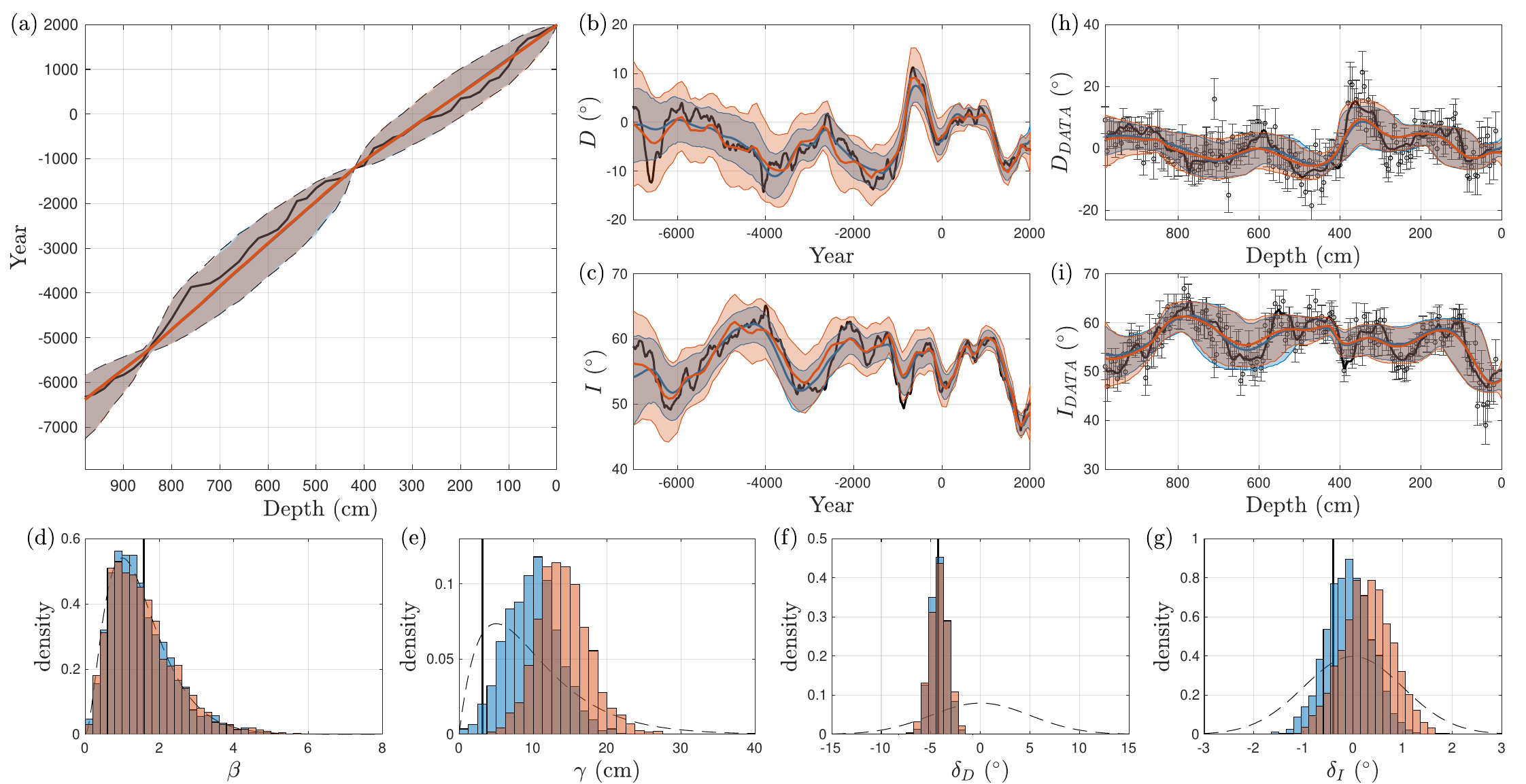}
\caption{Recovery of model parameters associated with the Lake Biwa (BIW) sediment record. CFF models based on Midpath prior (blue), Neutral\_top1 (red) are compared to the reference (black lines) and the prior distribution (dashed black lines), where appropriate. (a) Age-depth, (b-c) model predictions of $D(t)$ and $I(t)$ for the site coordinates, (d) lock-in shape $\beta$, (e) half lock-in depth $\gamma$, systematic errors in (f) declination $\delta_D$ and (g) inclination $\delta_I$ and (h-i) model-data comparison of $D_k$ and $I_k$. Due to the marginalization of the ages, the model predictions of $D_k$ and $I_k$ are not easy to illustrate. We therefore approximate them here by transferring the predicted $D(t)$ and $I(t)$, including pDRM effects and systematic errors, to depth-scale use random samples drawn from the prior age-depth model.}
\label{Fig_BIW_model_pred}
\end{figure*}

In Fig. \ref{Fig_GYL_model_pred}-\ref{Fig_BIW_model_pred}, we show model-data comparisons of three records from: (i) Gyltigesjön (GYL), Sweden, (ii) the Chilean margin (CHM) and (iii) Lake Biwa (BIW), Japan. For illustration purposes, we use random age-depth models drawn from the prior (a) to transfer predictions of $D(t)$ and $I(t)$ (b-c), including pDRM effects and systematic errors (d-g), to the depth domain of the data (h-i). 

\end{document}